\documentclass[10pt,a4paper]{article}
\usepackage[utf8]{inputenc}
\usepackage{cite}
\usepackage{amsmath}
\usepackage{amsfonts}
\usepackage{amssymb}
\usepackage{amsthm}
\usepackage[left=15mm, top=20mm, right=15mm, bottom=20mm, nohead]{geometry}
\usepackage{graphicx}
\usepackage{stmaryrd}
\usepackage{xcolor}
\usepackage{tikz}
\usepackage{subcaption}
\usetikzlibrary{patterns}
\usetikzlibrary{backgrounds,scopes}
\usetikzlibrary{calc}
\usetikzlibrary{intersections}
\usepackage{url}

\usepackage{hyperref}

\usepackage[all]{xy}

\newcommand{\beq}{\begin{equation}}
\newcommand{\eeq}{\end{equation}}

\makeatletter
\newcommand{\vast}{\bBigg@{4}}
\newcommand{\Vast}{\bBigg@{5}}
\makeatother

\newtheorem{theorem}{Theorem}[section]
\newtheorem{proposition}[theorem]{Proposition}
\newtheorem{corollary}[theorem]{Corollary}
\newtheorem{lemma}[theorem]{Lemma}

\theoremstyle{definition}
\newtheorem{definition}[theorem]{Definition}

\theoremstyle{remark}
\newtheorem{remark}[theorem]{Remark}

\newcommand{\ep}{\epsilon}

\newcommand{\be}{\beta}

\renewcommand{\appendix}[1]{
    \addtocounter{section}{1}
    \setcounter{equation}{0}
    \renewcommand{\thesection}{\Alph{section}}
    \section*{Appendix \thesection\protect\indent #1}
    \addcontentsline{toc}{section}{Appendix \thesection\ \ \ #1}
}
\newcommand\encadremath[1]{\vbox{\hrule\hbox{\vrule\kern8pt
\vbox{\kern8pt \hbox{$\displaystyle #1$}\kern8pt}
\kern8pt\vrule}\hrule}}
\def\enca#1{\vbox{\hrule\hbox{
\vrule\kern8pt\vbox{\kern8pt \hbox{$\displaystyle #1$}
\kern8pt} \kern8pt\vrule}\hrule}}

\newcommand\figureframex[3]{
\begin{figure}[bth]
\hrule\hbox{\vrule\kern8pt
\vbox{\kern8pt \vbox{
\begin{center}
{\mbox{\epsfxsize=#1.truecm\epsfbox{#2}}}
\end{center}
\caption{#3}
}\kern8pt}
\kern8pt\vrule}\hrule
\end{figure}
}
\newcommand\figureframey[3]{
\begin{figure}[bth]
\hrule\hbox{\vrule\kern8pt
\vbox{\kern8pt \vbox{
\begin{center}
{\mbox{\epsfysize=#1.truecm\epsfbox{#2}}}
\end{center}
\caption{#3}
}\kern8pt}
\kern8pt\vrule}\hrule
\end{figure}
}

\newcommand{\bea}{\begin{eqnarray}}
\newcommand{\eea}{\end{eqnarray}}

\renewcommand{\and}{{\qquad {\rm and} \qquad}}

\newcommand{\tr}{{\,\rm tr}\:}

\newcommand{\Pint}{{\int\kern -1.em -\kern-.25em}}

\renewcommand{\thesection}{\arabic{section}}

\makeatletter
\@addtoreset{equation}{section}
\makeatother
\newcommand{\brem}{\begin{remark}\rm\small}
\newcommand{\er}{\end{remark}}
\newcommand{\bt}{\begin{theorem}}
\newcommand{\et}{\end{theorem}}
\newcommand{\bd}{\begin{definition}}
\newcommand{\ed}{\end{definition}}
\newcommand{\bp}{\begin{proposition}}
\renewcommand{\ep}{\end{proposition}}
\newcommand{\bl}{\begin{lemma}}
\newcommand{\el}{\end{lemma}}
\newcommand{\bc}{\begin{corollary}}
\newcommand{\ec}{\end{corollary}}
\newcommand{\beaq}{\begin{eqnarray}}
\newcommand{\eeaq}{\end{eqnarray}}

\def\be{\begin{eqnarray}}
\def\ee{\end{eqnarray}}
\def\nn{\nonumber}

\def\tr{{\rm tr}\,}

\def\Sch{{\rm Schur}}\def\Sch{{\chi}}
\def\Mac{{\rm Mac}}

\title{\vspace{-1cm}{\Large {\bf
     Quantization of Harer-Zagier formulas
    }
    \date{}
    \author{
      {\bf A. Morozov$^{a,b,c}$}\thanks{morozov.itep@mail.ru},
      {\bf A. Popolitov$^{a,b,c}$}\thanks{popolit@gmail.com},
      {\bf Sh. Shakirov$^{b}$}\thanks{shakirov.work@gmail.com}}
}}

\begin{document}

\maketitle
\vspace{-4.8cm}

\begin{center}
	\hfill ITEP/TH-17/20 \\
	\hfill IITP/TH-12/20 \\
	\hfill MIPT/TH-11/20
\end{center}

\vspace{2.7cm}

\begin{center}
  $^a$ {\small {\it Institute for Theoretical and Experimental Physics, Moscow 117218, Russia}}\\
  $^b$ {\small {\it Institute for Information Transmission Problems, Moscow 127994, Russia}}\\
  $^c$ {\small {\it Moscow Institute of Physics and Technology, Dolgoprudny 141701, Russia }}
\end{center}

\vspace{0.5cm}

\begin{abstract}
  We derive the analogues of the Harer-Zagier formulas
  for single- and double-trace correlators
  in the q-deformed Hermitian Gaussian matrix model.
  This fully describes single-trace correlators
  and opens a road to $q$-deformations
  of important matrix models properties,
  such as genus expansion and Wick theorem.
\end{abstract}

{\section{Introduction}\label{sec:introduction}
  {
    Matrix models \cite{book:M-random-matrices,
      paper:MM-exact-equation-for-the-loop-average-in-multicolor-qcd,
      paper:IZ-the-planar-approximation-II,
      paper:K-the-appearance-of-matter-fields,
      paper:KM-induced-qcs-at-large-n}
    are now ubiquitous in mathematical and theoretical physics
    \footnote{For a review of various interconnections, applications and notation,
      relevant for the present paper, see
    \cite{paper:M-integrability-and-matrix-models,
      paper:M-matrix-models-as-integrable-systems,
      paper:M-quantum-deformations-of-tau-functions}.}.
    Reduction (or reformulation) of a problem in matrix model terms often leads to significant progress,
    be it in the domain of SUSY gauge theories
    \cite{
      paper:MMS-matrix-model-conjecture-for-exact-BS-periods,
      paper:MMS-conformal-blocks-as-df-integral-discriminants,
      paper:MMS-on-the-df-representation-of-toric,
      paper:MMS-towards-a-proof-of-agt-mm,
      paper:MMS-a-direct-proof-of-agt-at-beta-1,
      paper:MMS-proving-agt-as-hs-duality,
      paper:IO-method-of-generating-q-expansion-coefficients,
      paper:ZP-localization-review,
      paper:LPSZ-solving-q-virasoro-constraints,
      paper:CLPZ-exact-susy-wilson-loops-on-s3,
      paper:CLZ-on-matrix-models-and-their-q-deformations,
      paper:MNZ-5d-sym-and-ads7,
      paper:NZ-q-virasoro-constraints-in-matrix-models,
      paper:NNZ-q-virasoro-modular-double-and-3d-partition-functions},
    enumerative geometry
    \cite{
      paper:MM-virasoro-constraints-for-kh-partition-function,
      paper:AMMN-on-kp-integrable-hurwitz-functions,
      paper:KL-combinatorial-solutions-to-integrable-hierarchies,
      paper:DBOPS-combinatorics-of-loop-equations-for-branched-covers-on-sphere,
      paper:KPS-topological-recursion-for-monotone-orbifold,
      paper:DBKPS-cut-and-join-for-monotone,
      paper:K-recursion-for-masur-veech,
      paper:CKL-polynomial-graph-invariants-and-the-kp-hierarchy,
      paper:KZ-rationality-in-map-and-hypermap,
      paper:KZ-virasoro-constraints-and-topological-recursion-for-dessins,
      paper:DBKOSS-polynomiality-of-hurwitz-numbers-bm-conj,
      paper:A-matrix-model-for-stationary-sector-of-gw-of-p1,
      paper:ABT-refined-open-intersection-numbers-and-the-kp-mm,
      paper:A-open-intersection-numbers-and-free-fields,
      paper:DL-on-the-goulden-jackson-vakil-conjecture-for-double-hurwitz-numbers,
      paper:BDKLM-double-hurwitz-numbers-polynomiality-topological-recursion-and-intersection-theory},
    the theory of symmetric functions
    \cite{paper:AKMM-shiraishi-functor-and-non-kerov,
      paper:MM-on-hamiltonians-for-kerov-functions,
      paper:MM-kerov-functions-for-composite-representations,
      paper:MM-kerov-functions-revisited,
      paper:OP-the-equivariant-gw-theory-of-p1,
      paper:OP-gw-theory-hurwitz-theory-and-completed-cycles,
      paper:OP-gw-theory-hurwitz-numbers-and-matrix-models}
    or even the quantum computation
    \cite{
      paper:SKDMSK-robust-architecture-for-programmable-universal-unitaries,
      paper:KM-quantum-r-matrices-as-universal-qubit-gates}.

    An interesting built-in feature of matrix models is their genus expansion
    -- the natural splitting of any correlator into the contributions that can be thought of
    as  associated with Riemann surfaces of particular genera,
    endowed (colored) with certain extra data. From the QFT point of view this expansion
    is nothing but WKB (perturbative) expansion, and in the simplest case of Hermitian Gaussian matrix model
    (HGMM, see Section~\ref{sec:background} for a definition) it literally takes the form of the summation over
    fat (ribbon) graphs, each living on a particular Riemann surface.
  }

  For example, the average of $\tr (X^4)$ in HGMM is equal to

\begin{align}
\big\langle \ \tr (X^4) \ \big\rangle = 2 N^3 + N
\end{align}
  and the three summands are related to three fat graphs,
  two of which live on the sphere, and one on the torus:

\begin{figure}[!h]
  \begin{center}
  \includegraphics[width=0.75\textwidth]{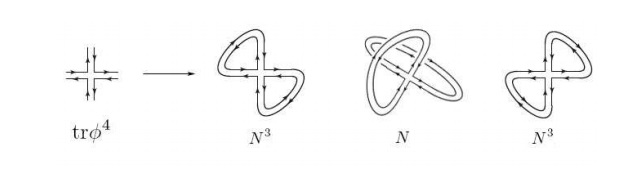}
  \end{center}
\end{figure}
  There are many ways to derive this formula in case of HGMM, for instance, using
  the Wick theorem, which in this case takes the form of gluing ribbons (MM propagator)
  to the discs with marked boundary points
  (for the details see, for instance, \cite{paper:AC-the-matrix-model-for-dessins-denfants,paper:DBOPS-combinatorics-of-loop-equations-for-branched-covers-on-sphere}).

  This simplicity, however, is lost in the case of q-deformed
  Hermitian Gaussian matrix model (qHGMM, see Section~\ref{sec:background}), where
  the corresponding average is equal to

\begin{align} \label{eq:trace-x4-quantum}
  \nn \left \langle \ \tr (X^4) \ \right \rangle_q = \ & \frac{1}{[2][4]}  \left( q^4 \dfrac{[N+3]!}{[N-1]!} + \dfrac{[N+2]!}{[N-2]!} - \dfrac{[N+1]!}{[N-3]!} + \dfrac{1}{q^4} \dfrac{[N]!}{[N-4]!} \right) = \emph{}
  \\ \nn \emph{} = \ & - \dfrac{(q^{-8}-q^{8})(q^{-6}-q^{6})}{(q^{-4}-q^{4})^2(q^{-2}-q^{2})(q^{-1}-q^{1})^2} q^{4N} +
  \\ \emph{} + \ & \dfrac{(q^{-6}-q^{6})(q^{-4}-q^{4})}{(q^{-3}-q^{3})(q^{-2}-q^{2})(q^{-1}-q^{1})^3} q^{2N} - \dfrac{(q^{-4}-q^{4})(q^{-3}-q^{3})}{(q^{-2}-q^{2})^2(q^{-1}-q^{1})^3} q^{2N} +
   \\ \nn \emph{} + \ &
   \dfrac{2}{(q^{-4}-q^{4})(q^{-1}-q^{1})^2} q^{-4N}
\end{align}
  Here we introduce $q$-numbers, $[n] = \frac{q^{-n}-q^{n}}{q^{-1}-q}$ and $q$-factorials, $[n]! = [1][2]\ldots[n]$. Not only does this formula feature strikingly new dependence on $N$ (in the form of $q^N$), the leading power of $q^N$ is no longer equal to the Euler characteristic of neither sphere nor torus.
The complexity persists for more complicated correlators and this makes fainting the hope
  to generalize Wick's theorem, genus expansion and other traditional matrix model
  structures beyond $q = 1$ case.

  However, the hope is revived by the main observation the present paper --
  explicit q-deformation of Harer-Zagier
  \cite{paper:HZ-the-euler-characteristic-of-the-moduli-space-of-curves}
  formula.

  This formula tells that Laplace transform in the variable $N$
  of the 1-point (single-trace) correlators can be explicitly presented as a
  fully-factorized rational function (see, also \cite{paper:MSh-exact-2-point-function-in-hmm}
  for a 2-point analog)
  \begin{align}
  \sum\limits_{N=0}^{\infty} \lambda^N \ \big\langle \ \tr (X^4) \ \big\rangle = \dfrac{3\lambda(1+\lambda)^2}{(1-\lambda)^4}
  \end{align}
  We find that its q-analogue is \textbf{not} any more complicated
  \begin{align}
  \sum\limits_{N=0}^{\infty} \lambda^N \ \big\langle \ \tr (X^4) \ \big\rangle_q = \dfrac{[3]_q\lambda q^2(q^2+\lambda)(q^4 + \lambda)}{(q^{4}-\lambda)(1-\lambda)(1-q^2\lambda)(1-q^4\lambda)}
  \end{align}
  where $[3]_q = q^2 + 1 + q^{-2}$ is a q-number.
  That is, $q$-deformation of Laplace transforms is  just a judicious insertion of $q$-monomials
  into some of the brackets of $q = 1$ answer. This is an example of q-generalization of Harer-Zagier formula,
  see Section~\ref{sec:harer-zagier-formulas} for the general case.

  The simplest way to derive the Harer-Zagier generating function (see Section~\ref{sec:hz-proof}),
  is by extensive use of the \emph{superintegrability} of the model
  -- an important property which takes form of exact solvability of correlators of Schur polynomials $\chi_{\lambda}$,
  \begin{align}
    \left\langle \chi_{\lambda}\{ X \} \ \right\rangle_q \sim \chi_{\lambda}\big( p_k^{*} \big)
  \end{align}
  where $p_k^{*}$ is a distinguished point in the space of arguments (power sums, or time-variables) of Schur polynomials.
  It is important to emphasize that superintegrability does persist in models where genus expansion and Wick theorem are not readily available.
  Study of Harer-Zagier functions can therefore shed light on potential generalizations and interplay of these structures.

  With help of this observation, it is straightforward to split the q-average
  of $\tr X^4$ into the contributions of different powers of $q^N$. Indeed, because of a general property of Fourier transform, expansion in powers of $q^N$ is essentially the expansion of the Harer-Zagier function in partial fractions:

  \begin{align}
    & \hspace{-22ex} \dfrac{[3]_q\lambda q^2(q^2+\lambda)(q^4 + \lambda)}{(q^{4}-\lambda)(1-\lambda)(1-q^2\lambda)(1-q^4\lambda)} = \dfrac{1}{1-q^4\lambda} \mbox{res}_{\lambda=q^{-4}}  \\ \nonumber + \dfrac{1}{1-q^2\lambda}\mbox{res}_{\lambda=q^{-2}}  & +  \dfrac{1}{1-\lambda} \mbox{res}_{\lambda=1} + \dfrac{q^4}{q^4-\lambda} \mbox{res}_{\lambda=q^{4}}
  \end{align}
  where in residues we recognize the familiar coefficients from \eqref{eq:trace-x4-quantum}
  \begin{align}
    \mbox{res}_{\lambda=q^{-4}} = & - \dfrac{(q^{-8}-q^{8})(q^{-6}-q^{6})}{(q^{-4}-q^{4})^2(q^{-2}-q^{2})(q^{-1}-q^{1})^2} \\ \notag
    \mbox{res}_{\lambda=q^{-2}} = & \dfrac{(q^{-6}-q^{6})(q^{-4}-q^{4})}{(q^{-3}-q^{3})(q^{-2}-q^{2})(q^{-1}-q^{1})^3} \\ \notag
    \mbox{res}_{\lambda=1} = & - \dfrac{(q^{-4}-q^{4})(q^{-3}-q^{3})}{(q^{-2}-q^{2})^2(q^{-1}-q^{1})^3} \\ \notag
    \mbox{res}_{\lambda=q^{4}} = & \dfrac{2}{(q^{-4}-q^{4})(q^{-1}-q^{1})^2}
  \end{align}
  We interpret this as follows: \textbf{the q-generalized Harer-Zagier provides
  a simple way to compute (1-point) correlators in qHGMM, by means of a partial fractions expansion.}
  This is the main practical result of the present paper, see
  Section~\ref{sec:harer-zagier-formulas}, Equations~\eqref{eq:explicit-residues1},\eqref{eq:explicit-residues2}.
  This result allows for efficient and rapid calculation of arbitrary 1-point correlators and thus, in particular, opens a way
  to make educated guesses about the shape of \textit{quantum spectral curve}
  \cite{
    paper:NS-quantization-of-integrable-systems-and-four-dimensional-gauge-theories,
    paper:MM-nekrasov-functions-and-exact-bs-integrals,
    paper:MM-nekrasov-functions-from-exact-bs-periods-sun,
    paper:MMM-on-agt-relations-with-surface-operator-insertion-and-a-stationary-limit-of-beta-ensembles,
    paper:NRS-darboux-coordinates-yy-functional-and-gauge-theory,
    paper:K-non-perturbative-quantum-geometry,
    paper:K-non-perturbative-quantum-geometry-ii,
    paper:GMM-wall-crossing-invariants-from-quantum-mechanics-to-knots,
    paper:MSS-the-qsc-for-double-hn,
    paper:DBMNPS-quantum-spectral-curve-for-gw-of-cp1,
    paper:GGM-non-perturbative-approaches-to-the-quantum-sw-curve,
    paper:EMR-resonances-and-pt-symmetry-in-quantum-curves}.

  In Sections~\ref{sec:multi-hz},\ref{sec:qt-generalization}~and~\ref{sec:q-genus-expansion}
  we outline possible further directions and applications, concentrating on problems and partial successes,
  but the actual development of these topics is postponed till future.
}

{\section{Background}\label{sec:background}
  %% ### vv Here we write the definition of matrix models and the fermionic (character expansion) formulas

\subsection{Hermitian Gaussian matrix model (HGMM)}

  HGMM can be defined as the following integral over the space of eigenvalues

  \begin{align}
    Z\{ p_k \} = \int\limits_{-\infty}^{\infty} \prod\limits_{i=1}^{N} e^{-x_i^2/2} \prod\limits_{i \neq j} (x_i - x_j) \ \exp\left( \sum\limits_{k} \dfrac{p_k}{k} (x_1^k + \ldots + x_N^k) \right) \ dx_1\ldots dx_N
  \end{align}
  \smallskip\\
  and admits a well-known exact solution in terms of Schur polynomials $\chi_R(p_k)$
  (here $R$ are Young diagrams and $p_k$-the time-variavbles),
  which can be also taken for the definition of the matrix model
  \cite{paper:AIMMVY-correspondence-between-feynman-diagrams,paper:IMM-complete-solution-to-gaussian-tensor-model,paper:IMM-from-kronecker-to-tableau}

  \begin{align}
    Z\{ p_k \} = \left\langle \exp\left( \sum\limits_{k}
    \dfrac{p_k}{k}\cdot (x_1^k + \ldots + x_N^k) \right) \ \right\rangle
   =\nn \\
   =\sum_R  \Sch_R(p_k)
   \Big\langle \Sch(x_1^k+\ldots +x_N^k)\Big\rangle
   = \sum\limits_R \dfrac{ \chi_R\big( \delta_{k,2} \big) }{ \chi_R\big( \delta_{k,1} \big) } \ \chi_R\big( p_k^{*} \big) \ \chi_R\big( p_k \big)
  \end{align}
  \smallskip\\
  where we applied the Cauchy formula.
  The special points in time-variables space are
  $p_k=\delta_{k,n} = \left\{\begin{array}{ccc} 1& & {\rm if} \ k =n \\ 0 && {\rm otherwise}\end{array}\right.$
  while
  \begin{align}
    p_k^{*} = N
  \end{align}
  is another distinguished point (the $q=1$ limit of topological locus, see below).
  This exact solution can be conveniently memorized as a character preservation property
  \begin{align}
    \notag \text{\textbf{The average of a character is again a character}} \\
    \left\langle \chi_R[ X ] \ \right\rangle = \left\langle\ \chi_R( x_1^k + \ldots + x_N^k ) \ \right\rangle = \dfrac{ \chi_R\big( \delta_{k,2} \big) }{ \chi_R\big( \delta_{k,1} \big) } \ \chi_R\big( p_k^{*} \big)
  \end{align}
  and by itself has deep and far going consequences
  \cite{paper:MM-on-the-complete-perturbative-solution-of-on-matrix-models,
    paper:MM-correlators-in-tensor-models-from-character-calculus,
    paper:IMM-complete-solution-to-gaussian-tensor-model,
    paper:CHPS-orbifolds-and-exact-solutions,
    paper:MPSh-on-qt-deformation-of-gaussian-mm,
    paper:IMM-tensorial-generalization-of-characters,
    paper:M-cauchy-formula-and-the-character-ring}.

  This apparent simplicity is obscured when looking at correlators of other symmetric quantities,
  different from Schur polynomials. Consider, for instance, the first few 1-point correlators, i.e. averages of power sums

  \begin{align}
    \left \langle \tr\{ X^2 \} \ \right\rangle = & \ \left\langle \ x_1^2 + \ldots + x_N^2 \ \right\rangle = N^2 \\ \notag
    \left\langle \tr\{ X^4 \} \ \right\rangle = & \ \left\langle \ x_1^4 + \ldots + x_N^4 \ \right\rangle = 2 N^3 + N \\ \notag
    \left\langle \tr\{ X^6 \} \ \right\rangle = & \ \left\langle \ x_1^4 + \ldots + x_N^4 \ \right\rangle = 5 N^4 + 10 N^2 \\ \notag
    & \dots
  \end{align}
  and so on. There is no closed expression for these correlators,
  in contrast with the Schur averages presented above.
  However, various generating functions exist -- and these are exactly the Harer-Zagier formulas.

\subsection{q-deformation of HGMM}

  The Hermitian Gaussian matrix model admits
  a $q$-generalization

  \begin{align}
    Z_{q}\{ p_k \} = \int\limits_{-\nu}^{\nu} \prod\limits_{i=1}^{N} \rho(x_i) \ \prod\limits_{i \neq j} (x_i - x_j) \ \exp\left( \sum\limits_{k} \dfrac{p_k}{k} (x_1^k + \ldots + x_N^k) \right) \ dx_1\ldots dx_N
  \end{align}
  \smallskip\\
  where $\nu = (1-q)^{-1/2}$ and $\rho(x)$ is a $q$-deformation of the Gaussian distribution
  and the integral $\int_{-\nu}^\nu dx$ is, in fact, the $q$-deformed (so-called, Jackson) integral
  (for details see \cite{paper:LPSZ-solving-q-virasoro-constraints,paper:MPSh-on-qt-deformation-of-gaussian-mm}).
  Similarly to the original model there is a nice solution in terms of Schur functions:

  \begin{align}
    Z_q\{ p_k \} = \left\langle \exp\left( \sum\limits_{k} \dfrac{p_k}{k} (x_1^k + \ldots + x_N^k) \right) \ \right\rangle_q = \sum\limits_{R} \dfrac{ \chi_R\big( \delta_{k|2} \big) }{ \chi_R\big( \delta_{k|1} \big) } \ \chi_R\big( p_k^{*} \big) \ \chi_R\big( p_k \big)
  \end{align}
  \smallskip\\
  where points $\delta_{k|n} = n \frac{(q^{-1}-q)^{k/n}}{q^{-k}-q^k}$ if $k$ is divisible by $n$ and 0 otherwise, and
  \begin{align}
    p_k^{*} = \dfrac{q^{-Nk}-q^{Nk}}{q^{-k} - q^{k}}
  \end{align}
  is an important distinguished point
  (the so-called \emph{topological locus}
  that plays an important role in knot theory
  \cite{paper:BDGMMMRSS-distinguishing-mutant-knots,paper:MST-a-novel-symmetry-of-colored-homfly,paper:BDGMMMRSS-difference-of-mutant-knot-invariants,paper:MST-a-new-symmetry-of-colored-alexander}). In the language of correlators, a correlator of a Schur function is nicely expressed in terms of Schur functions again:

  \begin{align}\label{eq:q-schur-average}
    \left\langle \chi_R\{ X \} \ \right\rangle_q = \left\langle\ \chi_R( x_1^k + \ldots + x_N^k ) \ \right\rangle_q = \dfrac{ \chi_R\big( \delta_{k|2} \big) }{ \chi_R\big( \delta_{k|1} \big) } \ \chi_R\big( p_k^{*} \big)
  \end{align}
%  \smallskip\\
%  Just like in the original model, the 1-point correlators are nowhere near simple.

  {\subsection{The triad of definitions} \label{sec:def-triade}

    In this subsection we do not pay so much attention to the difference
    between usual $\scriptstyle (1 - q^n)/(1-q)$
    and symmetric $\scriptstyle (q^n - q^{-n})/(q - q^{-1})$ definitions of the $q$-numbers.

    In fact, the definitions can be lifted one level up, to the $(q,t)$-deformed eigenvalue model.
    There are three possible ways to define the $(q,t)$-deformation of the Hermitian Gaussian matrix model (qtHGMM)

    The first one is with help of explicit Jackson integration (\cite{paper:LPSZ-solving-q-virasoro-constraints}, Equation~(2.6))
    \begin{align}\label{eq:qt-hgmm-jackson-definition}
      Z_{(q,t)}\{p_k\} = \int_{-\nu}^\nu\prod_{i=1}^N d_q x_i \prod_{j=1}^N(x_j^2 q^2 \nu^{-2}; q^2)_\infty
      \prod_{k\neq l} x_l^\beta \dfrac{(x_k/x_l;q)_\infty}{(t x_k/x_l; q)_\infty} \exp\left( \sum\limits_{k} \dfrac{p_k}{k} (x_1^k + \ldots + x_N^k) \right),
    \end{align}
    where, as usual, $t = q^\beta$, $\nu = 1/\sqrt{1 - q}$, $(x ; q)_\infty = \prod_{m = 0}^\infty (1 - x q^m)$ is the $q$-Pochhammer symbol
    and $\int_{-\nu}^\nu d_q x$ is the $q$-Jackson integral
    \begin{align}
      \int_{-\nu}^\nu d_q x f(x) := (1 - q) \sum_{n=0}^\infty \nu q^n \left [ g(\nu q^n) + g(-\nu q^n)\right ]
    \end{align}
    While this definition may seem \textit{ad hoc}, in fact the Jackson integral can be reinterpreted as a sum over poles
    of some meromorphic function, after which $Z_{(q,t)}\{p_k\}$ becomes nothing but the $\mathcal{N}=2$
    SUSY Yang-Mills-Chern-Simons partition function on $D_2 \times S^1$.

    This definition can be used to directly evaluate any trace product correlator only at integer values of $\beta$,
    nevertheless, it is useful for cross-checks.

    A non-trivial corollary of \ref{eq:qt-hgmm-jackson-definition} is
    that $Z\{p_k\}$ satisfies the system of $(q,t)$-deformed Virasoro constraints \cite{paper:LPSZ-solving-q-virasoro-constraints},
    Equation~(4.20). Moreover, by clever choice of the equations from the system one can obtain
    a decomposition formula for any trace product correlator \cite{paper:LPSZ-solving-q-virasoro-constraints}, Equation~(5.4),
    which is a bit lengthy to provide it here.
    Supplemented with initial conditions $C_\emptyset = 1$ and $C_1 = 0$ this decomposition formula
    can be used as an alternative definition for (normalized) correlators of qtHGMM.

    \subsection{On $q,t$-deformed HGMM}

  The $q$-deformation can be further promoted to a $q,t$-analogue of the Gaussian matrix model
  ($\beta$-ensemble),
  where the measure $\prod(x_i-x_j)$ is $\beta$-deformed with $t=q^{\beta}$,
  and Schur functions deform to Macdonald functions. 
  We consider this far-going generalization
    only briefly in this paper, see also Section~\ref{sec:qt-generalization}.

    As with other Gaussian models, 
    the most practical definition of qtHGMM is by means of the $(q,t)$-character expansion property 
    ("superintegrability")
    \cite{paper:IMM-complete-solution-to-gaussian-tensor-model,
      paper:IMM-from-kronecker-to-tableau,
      paper:MM-on-the-complete-perturbative-solution-of-on-matrix-models,
      paper:MM-correlators-in-tensor-models-from-character-calculus}
    -- the approach taken in \cite{paper:MPSh-on-qt-deformation-of-gaussian-mm}
    \begin{align}
      \left\langle \Mac_R \right \rangle_{q,t} = \Mac_R\left(\delta_{k|2}^{(qt)}\right)
      \frac{\Mac_R(\pi^*)}{\Mac_R\left(\delta_{k|1}^{(qt)}\right)},
    \end{align}
    where the $\pi^*$ is the $(q,t)$ topological locus
    \begin{align}
      \pi^*: \ \ p_k = \dfrac{t^{-Nk} - t^{Nk}}{t^{-k} - t^k}
    \end{align}
    and the ``special'' loci $\delta_{k|1}^{(qt)}$ and $\delta_{k|2}^{(qt)}$ are defined by
    \begin{align} \label{eq:special-qt-loci}
      \delta_{k|n}^{(qt)} = n \dfrac{(q^{-1} - q)^{k/n}}{t^{-k}-t^k} \cdot
      \left \{
      \begin{array}{l}
        1 \text{ if $k$ is divisible by $n$} \\
        0 \text{ otherwise}
      \end{array}
      \right .
    \end{align}
    This choice is somewhat {\it ad hoc} -- 
    it is partly justified in \cite{paper:MPSh-on-qt-deformation-of-gaussian-mm},
    but can be easily changed to fit other interesting cases, up to 
    qtHGMM, inspired by arbitrary knot superpolynomials in the spirit of
    \cite{paper:MM-sum-rules-for-characters-from-character-preservation-property}.
    In the present paper we limit ourselves to \eqref{eq:special-qt-loci}.

    The relation between the three definitions (integral, Virasoro and Macdonald) is as follows.
    Ref.\cite{paper:LPSZ-solving-q-virasoro-constraints} provides strict derivation
    of the Virasoro definition from the eigenvalue-integral definition \eqref{eq:qt-hgmm-jackson-definition}.
    However, there is no way
    to restore the partition function at $p_k = 0$ from the Virasoro definition.
    The coincidence of Macdonald definition with the first two is a long-known experimental fact,
    but currently proving it directly from the definitions
    remains an open problem.
    Note, however, that the analogous equivalence is proved for the Selberg integral
    in the Macdonald book \cite{book:M-symmetric-functions-and-hall-polynomials}, Chapter~VI, Section~9, Example~3.
  }
}

{\section{Harer-Zagier formulas}\label{sec:harer-zagier-formulas}
  %% ### vv Here we write our main observation about the q-Harer-Zagier formulas

  The initial Harer-Zagier statement\cite{paper:HZ-the-euler-characteristic-of-the-moduli-space-of-curves}
  is, that the generating function for HGMM one-point correlators,
  normalized by a peculiar double-factorial factors, is an elementary function
  \begin{align}
    \sum_{k = 0}^\infty \left \langle \tr X^{2k} \right \rangle \frac{x^{2k}}{(2 k - 1)!!}= & \
    \frac{1}{2 x^2}
    \left(\left(\frac{1 + x^2}{1 - x^2}\right)^N - 1 \right)
  \end{align}

  Subjected to additional Laplace transform in $N$, this function becomes even simpler
  \begin{align}
    \sum_{N=0}^\infty \lambda^N \sum_{k = 0}^\infty \left \langle \tr X^{2k} \right \rangle
    \frac{x^{2k}}{(2 k - 1)!!} = & \
    \frac{\lambda}{(1 - \lambda)} \frac{1}{(1 - \lambda) - (1 + \lambda) x^2}
  \end{align}
  so that, in fact, its expansion coefficients in $x$ -- the Laplace transforms of individual
  one-point correlators -- are also very simple expressions.
  Explicitly, the first few correlators, summed over N with weight $\lambda^N$ give
  \begin{align}
    \sum_{N = 0}^{\infty} \lambda^N \left\langle \tr\{ X^2 \} \ \right\rangle = & \ \sum_{N = 0}^{\infty} N^2 \lambda^N = \dfrac{\lambda(1+\lambda)}{(1-\lambda)^3} \\ \notag
    \sum_{N = 0}^{\infty} \lambda^N \left\langle \tr\{ X^4 \} \ \right\rangle = & \ \sum_{N = 0}^{\infty} ( 2 N^3 + N ) \lambda^N = 3 \dfrac{\lambda(1+\lambda)^2}{(1-\lambda)^4} \\ \notag
    \sum_{N = 0}^{\infty} \lambda^N \left\langle \tr\{ X^6 \} \ \right\rangle = & \ \sum_{N = 0}^{\infty} ( 5 N^4 + 10 N^2 ) \lambda^N = 15 \dfrac{\lambda(1+\lambda)^3}{(1-\lambda)^5}
    \\ & \ \dots \notag
  \end{align}
  It is straightforward to imply (and then prove, for instance, with help of Toda equation)
  the general formula for Laplace transform of 1-point correlator:
  \begin{align}\label{eq:hz-q-equals-1}
    \boxed{ \ \ \ \sum_{N = 0}^{\infty} \lambda^N \left\langle \tr\{ X^{2m} \} \ \right\rangle = (2m-1)!! \dfrac{\lambda(1+\lambda)^{m}}{(1-\lambda)^{m+2}} \ \ \ }
  \end{align}
  \smallskip\\
  These kind of formulas we also call Harer-Zagier formulas.

  The surprising observation of the present paper is that q-analogues of Harer-Zagier formulas are equally simple,

  \begin{align}
    \sum_{N = 0}^{\infty} \lambda^N \left\langle \tr\{ X^2 \} \ \right\rangle_q = & \ \dfrac{q\lambda(q^2+\lambda)}{(q^2-\lambda)(1-\lambda)(1-q^2\lambda)}
    \\ \notag
    \sum_{N = 0}^{\infty} \lambda^N \left\langle \tr\{ X^4 \} \ \right\rangle_q = & \ [3] \dfrac{q^2\lambda(q^2+\lambda)(q^4+q\lambda)}{(q^4-\lambda)(1-\lambda)(1-q^2\lambda)(1-q^4\lambda)}
    \\ \notag
    \sum_{N = 0}^{\infty} \lambda^N \left\langle \tr\{ X^6 \} \ \right\rangle_q = & \ [3][5] \dfrac{q^3\lambda(q^2+\lambda)(q^4+\lambda)(q^6+\lambda)}{(q^6-\lambda)(1-\lambda)(1-q^2\lambda)(1-q^4\lambda)(1-q^6\lambda)}
  \end{align}
  \smallskip\\
  and so on. Generally

  \begin{align} \label{eq:q-harer-zagier-concise}
    \boxed{ \ \ \ \sum_{N = 0}^{\infty} \lambda^N \left\langle \tr\{ X^{2m} \} \ \right\rangle_q = [2m-1]!! \ \dfrac{q^m \lambda\prod_{n=1}^{m} (q^{2n} + \lambda)}{(q^{2m}-\lambda)\prod_{n=0}^{m}(1-q^{2n}\lambda)} \ \ \ }
  \end{align}
  \smallskip\\
  where $[2m-1]!! = [1][3][5]\ldots[2m-1]$ is the $q$-double factorial.
  Using this result, it is straightforward to derive a simple formula for the 1-point correlators:

  \begin{align} \label{eq:explicit-residues1}
    \left\langle \tr\{ X^{2m} \} \ \right\rangle_q
    = q^{-2 N m} \mbox{res}_{\lambda = q^{2m}} + \sum\limits_{a = 0}^{m} \ q^{2 N a} \ \mbox{res}_{\lambda = q^{-2a}} \
  \end{align}
  \smallskip\\
  by means of evaluating the contributions of all the residues
  \begin{align} \label{eq:explicit-residues2}
    \mbox{res}_{\lambda = q^{-2a}} = & \ [2m-1]!! \ q^{m-2 a} \ \dfrac{\prod_{n=1}^{m} (q^{2n}+q^{-2a})}{(q^{2m}-q^{-2a})\mathop{\prod_{n=0}^{m}}\limits_{n \neq a}(1-q^{2n-2a})} \\ \notag
    \mbox{res}_{\lambda = q^{2m}} = & \ [2m-1]!! \ q^m \ \dfrac{\prod_{n=1}^{m} (q^{2n} + q^{2m})}{\prod_{n=0}^{m}(1-q^{2n+2m})}
  \end{align}
  \smallskip\\

  Note that such a formula was \textit{impossible} on the $q = 1$ level. There, in \eqref{eq:hz-q-equals-1}, all the simple poles collided
  at $\lambda = 1$ and there was no way to isolate their individual contributions and get the simple factorized expressions.
  Moreover, each of the residues \eqref{eq:explicit-residues2} is singular at $q = 1$, while $N$-dependence of each separate summand
  in \eqref{eq:explicit-residues1} vanishes at $q = 1$. It is the competition between these two effects that
  restores the well-known polynomial dependence on $N$ at $q = 1$.

}

{\section{Derivation of Harer-Zagier formulas}\label{sec:hz-proof}

  To derive the q-generalized Harer-Zagier formula, we use the following well-known expansion of power sums in Schur polynomials,
  \begin{align}
    \tr X^{2m} = \sum\limits_{\ell = 0}^{2m-1} \ (-1)^{\ell} \ \chi_{[2m-\ell,1^\ell]}[X]
  \end{align}
  \smallskip\\
  where $1^{\ell} = \underbrace{1,\ldots,1}_{\ell \mbox{ times}}$. Therefore,

  \be \label{stexpan}
    \big\langle \tr X^{2m} \big\rangle_q = \sum\limits_{\ell = 0}^{2m-1} \ (-1)^{\ell} \ \big\langle \chi_{[2m-\ell,1^\ell]}[X]\big\rangle_q
  \ee
  \smallskip\\
  Using the explicit expression for the correlator of a Schur polynomial~\eqref{eq:q-schur-average}, we find

  \begin{align}
    \big\langle \chi_{[2m-\ell,1^\ell]}[X]\big\rangle_q = &
    (-1)^{\ell+\ell_{even}/2} \ q^{-\ell_{even}^2/4+(2m-\ell_{even})^2/4}
    \ [2m - \ell_{even} - 1]!!
    \ [\ell_{odd}]!! \times \emph{} \\ \nonumber &
    \emph{} \times \dfrac{[N]}{[2m]}
    \prod\limits_{i = 2}^{\ell+1} \dfrac{[N+1-i]}{[1-i]}
    \prod\limits_{j = 2}^{2m-\ell} \dfrac{[N+j-1]}{[j-1]}
  \end{align}
  \smallskip\\
  where $\ell_{even} = 2 \lceil\frac{\ell}{2}\rceil$ and $\ell_{odd} = \ell_{even} - 1$. This can be rewritten as

  \begin{align}
    \nonumber \big\langle \chi_{[2m-\ell,1^\ell]}[X]\big\rangle_q = &
    (-1)^{\ell_{even}/2} \ q^{-\ell_{even}^2/4+(2m-\ell_{even})^2/4}
    \ [2m - \ell_{even} - 1]!!
    \ [\ell_{odd}]!! \times \emph{} \\ &
    \emph{} \times \dfrac{1}{[2m]} \dfrac{1}{[2m-\ell-1]![\ell]!}
    \dfrac{[N+2m-\ell-1]!}{[N-\ell-1]!}
  \end{align}
  \smallskip\\
  Summing over $N$, we find

  \begin{align}
    \nonumber \sum\limits_{N = 0}^{\infty} \ \lambda^N \ \big\langle \tr X^{2m} \big\rangle_q & =
    \sum\limits_{\ell = 0}^{2m-1} \ (-1)^{\ell+\ell_{even}/2} \ q^{-\ell_{even}^2/4+(2m-\ell_{even})^2/4}
    \ [2m - \ell_{even} - 1]!!
    \ [\ell_{odd}]!! \times \emph{} \\ &
    \emph{} \times \dfrac{1}{[2m]} \dfrac{1}{[2m-\ell-1]![\ell]!}
    \sum\limits_{N = 0}^{\infty} \ \lambda^N \ \dfrac{[N+2m-\ell-1]!}{[N-\ell-1]!} =
    \\ \notag
    = &
    \sum\limits_{\ell = 0}^{2m-1} \ (-1)^{\ell+\ell_{even/2}} \ q^{-\ell_{even}^2/4+(2m-\ell_{even})^2/4}
    \ [2m - \ell_{even} - 1]!!
    \ [\ell_{odd}]!! \times \emph{}
    \\ \notag &
    \emph{} \times [2m-1]! \dfrac{\lambda^{\ell+1} }{[2m-\ell-1]![\ell]!}
    \ \prod\limits_{k = -m}^{m} \dfrac{1}{1-q^{2k} \lambda} =
    \\ \notag
    =
    \prod\limits_{k = -m}^{m} \dfrac{1}{1-q^k \lambda} \sum\limits_{\ell = 0}^{2m-1} & \ (-1)^{\ell+\ell_{even/2}} \ q^{-\ell_{even}^2/4+(2m-\ell_{even})^2/4} \ \dfrac{[2m - \ell_{even} - 1]!! [\ell_{odd}]!! [2m-1]!}{[2m-\ell-1]![\ell]!} \lambda^{\ell+1}
  \end{align}
  \smallskip\\
  To complete the derivation, we note the following algebraic identity:
  \begin{align}
    \nonumber \sum\limits_{\ell = 0}^{2m-1} \ (-1)^{\ell+\ell_{even/2}} \ q^{-\ell_{even}^2/4+(2m-\ell_{even})^2/4} \ \dfrac{[2m - \ell_{even} - 1]!! [\ell_{odd}]!! [2m-1]!}{[2m-\ell-1]![\ell]!} \lambda^{\ell+1} = \emph{} \\
    \hspace{10ex} \emph{} \boxed{ = [2m-1]!! \ q^{m^2} \ \dfrac{\lambda}{q^{2m}-\lambda} \ \prod\limits_{n=1}^{m} (q^{4n} - \lambda^2) },
  \end{align}
  which is a relative of $q$-Newton's binomial formula
  (see, for instance, \cite{book:KC-quantum-calculus}).

  Putting everything together, we obtain our main result

  \begin{align}
    \boxed{ \ \ \ \sum_{N = 0}^{\infty} \lambda^N \left\langle \ \tr  X^{2m}  \ \right\rangle_q =
     (2m-1)!!_q \ \dfrac{q^m \lambda\prod_{n=1}^{m} (q^{2n} + \lambda)}{(q^{2m}-\lambda)\prod_{n=0}^{m}(1-q^{2n}\lambda)} \ \ \ }
  \end{align}
  \smallskip\\
  which is the q-generalization of Harer-Zagier formula.
}

{\section{Towards multi-point Harer-Zagier formulas}\label{sec:multi-hz}

  We begin with reminding some results from \cite{paper:MSh-exact-2-point-function-in-hmm}
  at $q=1$.
  The $n$-point Harer-Zagier generating functions are made from {\it irreducible} correlators
  $K_{i_1,\ldots,i_n}(N)$
  \begin{align}
    \rho_{HZ}(s_1,\dots,s_n) = \sum_{N=0}^\infty \lambda^N
    \sum_{i_1 \dots i_n=0}^\infty K_{i_1,\ldots, i_n}(N)\cdot
    \frac{s_1^{i_1} \cdot \ldots \cdot s_n^{i_n}}{{\rm double\ factorials}}
  \end{align}
  and
  %at least 1-,2- and 3-point functions
  they always are elementary functions.
  For instance, the 2-point functions are given by
  \begin{align} \label{rho}
   % \rho_{HZ}(s_1, s_2) = & \ \rho_{HZ}^{(odd)}(s_1, s_2) + \rho_{HZ}^{(even)}(s_1, s_2)
    %\\ \notag
    \rho_{HZ}^{(odd)}(s_1, s_2) =
    \sum_{N=0}^\infty \lambda^N
    \sum_{i_1, i_2=1 }^\infty  K_{2i_1-1,2i_2-1}(N)\cdot
    \frac{s_1^{2i_1-1}s_2^{2i_2-1}}{(2i_1-1)!!(2i_2-1)!!}=
     & \ \frac{\lambda}{(1 - \lambda)^{3/2}}
    \frac{\arctan \left (
      \frac{s_1 s_2 \sqrt{\lambda - 1}}{\sqrt{\lambda - 1 + (\lambda + 1)(s_1^2 + s_2^2)}}
      \right )}{\sqrt{\lambda - 1 + (\lambda + 1)(s_1^2 + s_2^2)}}
    \\ \notag
    \rho_{HZ}^{(even)}(s_1, s_2) =
    \sum_{N=0}^\infty \lambda^N
    \sum_{i_1, i_2 =1}^\infty K_{2i_1,2i_2}(N)\cdot
    \frac{s_1^{2i_1}s_2^{2i_2}}{(2i_1-1)!!(2i_2-1)!!} =
    & \ \frac{s_1 s_2}{s_1^2 - s_2^2}
    \left (
    s_1 \frac{\partial}{\partial s_1} - s_2 \frac{\partial}{\partial s_2}
    \right ) \rho_{HZ}^{(odd)}(s_1, s_2)
  \end{align}

  Extension to $q\neq 1$ is rather straightforward, for example the simplest odd correlators are:
\be
\begin{array}{cl}
\sum_{N=0}^\infty \lambda^N \left \langle \tr X^{2m-1}\, \tr X \right \rangle_q = &\!\!\!\!
[2m-1]!!\cdot \frac{q^m\lambda\prod_{i=1}^{m-1}(q^{2i}+\lambda)}{\prod_{i=0}^m(1-q^{2i}\lambda)}
\\ \\
\sum_{N=0}^\infty \lambda^N \left \langle \tr X^{2m-1}\, \tr X^3 \right \rangle_q = &\!\!\!\!
[2m-1]!!\cdot \frac{q^{m+1}\lambda\prod_{i=1}^{m-2}(q^{2i}+\lambda)}{\prod_{i=0}^{m+1}(1-q^{2i}\lambda)}
\Big\{[2m+1]\Big(\lambda^2+q^{2m-1}[2]\lambda+q^{4m-2}\Big) +2q^{2m-1}[2m-2]\lambda\Big\}
%\\ \\
%\sum_{N=0}^\infty \lambda^N \left \langle \tr X^{2m-1}\, \tr X^5 \right \rangle_q = &\!\!\!\!
%[2m-1]!!\cdot \frac{q^{m+2}\lambda\prod_{i=1}^{m-3}(q^{2i}+\lambda)}{\prod_{i=0}^{m+2}(1-q^{2i}\lambda)}
%\cdot  \\
%&\cdot\Big\{[2m+3][2m+1]\Big(\lambda^4+q^{2m-1}[4]\lambda^3+q^{4m-2}\frac{[4][3]}{[2]}\lambda^2
%+ q^{6m-3}[4]\lambda+q^{8m-4}\Big) +   \\
%& \!\!\!\!\!\!\!\!\!\!\!\!\!\!\!\!\!\!\!\!\!\!\!\!
%+2q^{2m-1}[2m-2]\lambda\Big([2m+1](\lambda^2+q^{4m-2})
%+[5]q^{2m}(\lambda^2+q^{-2})\Big)
%+2q^{4m-2}[2m][2m-2][5]\lambda^2\Big\}
\\ \\
\ldots
\nn
\end{array}
\ee
i.e. for $m_1\geq m_2$
\be \label{odd2corr}
\sum_{N=0}^\infty \lambda^N \left \langle \tr X^{2m_1-1}\, \tr X^{2m_2-1} \right \rangle_q =  
[2m_1-1]!!\cdot \frac{q^{m_1+m_2-1}\lambda\prod_{i=1}^{m_1-m_2}(q^{2i}+\lambda)}
{\prod_{i=0}^{m_1+m_2-1}(1-q^{2i}\lambda)}
%\cdot\nn\\
\cdot{\rm Pol}_{2m_2-2}(m_1|\lambda) 
\nn \\
 = [2m_1+2m_2-1]!!\cdot \frac{q^{m_1+m_2-1}\lambda\prod_{i=1}^{m_1+m_2-1}(q^{2i}+\lambda)}{\prod_{i=0}^{m_1+m_2-2}(1-q^{2i}\lambda)}
+ \ldots
%\nn \\
%& [2m_1-1]!!\cdot %\frac{q^{m_1+m_2-1}\lambda\prod_{i=1}^{m_1-m_2}(q^{2i}+\lambda)}{\prod_{i=0}^{m_1+m_2-1}(1-q^{2i}\lambda)
\ee
while the simplest even ones are
\be \label{evencorr2}
\sum_{N=0}^\infty \lambda^N \left \langle \left \langle\tr X^{2m}\, \tr X^2 
\right \rangle\right \rangle_q   
:= \sum_{N=0}^\infty \lambda^N \left(\left \langle  \tr X^{2m}\, \tr X^2 \right \rangle_q 
 - \left \langle  \tr X^{2m}\  \right \rangle_q
 \left \langle  \ \tr X^2 \  \right \rangle_q    \right) \ 
= \nn \\ 
= [2m] [2m-1]!!  \frac{q^{m}\lambda}{(q^{2m}+\lambda  )}
          \frac{\prod_{i=1}^{m+1}(q^{2i}+\lambda )}{\prod_{i=0}^{m+1}(1 - q^{2i} \lambda)}
\ \ = \ \ \frac{[2m]}{q\cdot[2m+1]}\cdot \frac{q^{2m+2}+\lambda}{q^{2m}+\lambda} \cdot  
\sum_{N=0}^\infty \lambda^N \left \langle  \tr X^{2m+1}\, \tr X \ 
\right \rangle_q             
\ee
\be \label{evencorr4}
\sum_{N=0}^\infty \lambda^N \left \langle \left \langle\tr X^{2m}\, \tr X^4
\right \rangle\right \rangle_q
:= \sum_{N=0}^\infty \lambda^N \left(\left \langle  \tr X^{2m}\, \tr X^4 \right \rangle_q
 - \left \langle  \tr X^{2m}\  \right \rangle_q
 \left \langle  \ \tr X^4 \  \right \rangle_q    \right) \
= \nn \\ 
= \ \  \frac{ (q^{2m+4}+\lambda)}{(q^{2m-2}+\lambda)}\cdot\left(
\frac{[2m-2]}{q^3\cdot[2m+1]}  \cdot
\sum_{N=0}^\infty \lambda^N \left \langle  \tr X^{2m+1}\, \tr X^3 \
\right \rangle_q 
+\right. \nn \\ \left. 
+\frac{[3][2m+2]}{q\cdot[2m+1][2m+3]}\cdot \frac{(1+q^{2m-2}\lambda) }
{(q^{2m+2}+\lambda)} \cdot
\sum_{N=0}^\infty \lambda^N \left \langle  \tr X^{2m+3}\, \tr X \
\right \rangle_q \right)
\ee
%\be \label{evencorr6}
%\sum_{N=0}^\infty \lambda^N \left \langle \left \langle\tr X^{2m}\, \tr X^6
%\right \rangle\right \rangle_q
%:= \sum_{N=0}^\infty \lambda^N \left(\left \langle  \tr X^{2m}\, \tr X^6 \right \rangle_q
% - \left \langle  \tr X^{2m}\  \right \rangle_q
% \left \langle  \ \tr X^6 \  \right \rangle_q    \right) \
%= \nn \\
%= \ \ \frac{q^{2m+6}+\lambda}{q^{2m-4}+\lambda} \cdot\left(
%\frac{[2m-4]}{q^5\cdot[2m+1]}\cdot 
%\sum_{N=0}^\infty \lambda^N \left \langle  \tr X^{2m+1}\, \tr X^5 \
%\right \rangle_q 
%+\right. \nn \\ \left.
%+ \frac{[5][2m]}{q^3\cdot[2m+1][2m+3]}\cdot \frac{(1+q^{2m-4}\lambda) }
%{(q^{2m}+\lambda) } \cdot
%\sum_{N=0}^\infty \lambda^N \left \langle  \tr X^{2m+3}\, \tr X \
%\right \rangle_q
%+\right.\nn \\ \left.
%+ \frac{[5][3][2m+2]}{q\cdot[2m+1][2m+3]}\cdot \frac{(1+q^{2m}\lambda)(1+q^{2m-4}\lambda)}
%{(q^{2m}+\lambda)(q^{2m+4}+\lambda)} \cdot
%\sum_{N=0}^\infty \lambda^N \left \langle  \tr X^{2m+3}\, \tr X \
%\right \rangle_q\right)
%\ee
They are natural $q$-deformations of (\ref{rho}).

\bigskip

We now sketch the steps, leading to $q$-deformed formulas of this type.
As a generalization of (\ref{stexpan}) the full correlators
\be
C_\Delta = C_{i_1,\ldots,i_n} :=\left< \tr X^{i_1} \ldots \tr X^{i_n}\right>_q
\ee
for a Young diagram $\Delta=(i_1\geq i_2\geq\ldots\geq i_n)$ are equal to
\be
C_\Delta = \sum_{R\vdash \Delta} {\rm Chi}_{R,\Delta}\,\cdot <\Sch_R>_q =
\sum_{R\vdash \Delta} {\rm Chi}_{R,\Delta}\cdot \frac{\Sch_R(\delta_{k|2})}{\Sch_R(\delta_{k|1})}
\cdot\Sch_R(p_k^*)
\ee
where ${\rm Chi}_{R,\Delta}$ are easily available symmetric-group characters.\footnote{
  These characters satisfy orthogonality conditions
  $$
  \sum_{\Delta\vdash|R|} \frac{{\rm Chi}_{R,\Delta}{\rm Chi}_{R',\Delta}}{z_\Delta} = \delta_{R,R'}
  \ \ \Longleftrightarrow \ \  \sum_{R\vdash|Delta|}  {\rm Chi}_{R,\Delta}{\rm Chi}_{R,\Delta'}
  = {z_\Delta} = \delta_{\Delta,\Delta'}
  $$
  and appear in expansions of Schur polynomials
  $$
  \Sch_R(p) = \sum_{\Delta\vdash |R|} \frac{{\rm Chi}_{R,\Delta} p^\Delta}{z_\Delta}
  \ \ \Longrightarrow \ \ p^\Delta := \prod_{i=1}^{l_\Delta} p_{\Delta_i} =
  \sum_{R\vdash |\Delta|} {\rm Chi}_{R,\Delta}\Sch_R(p)
  $$
  They are related to characters $\varphi(R,\Delta)$ in
  \cite{paper:MMN-complete-set-of-cut-and-join-operators-in-the-hk-theory}
  by rescaling
  ${\rm Chi}_{R,\Delta} = z_\Delta\cdot \Sch_R(\delta_{k,1})\cdot \varphi(R,\Delta)$.
}
Irreducible correlators arise after subtractions, e.g. $K_{i_1i_2}:=C_{i_1i_2} -C_{i_1}C_{i_2}$
or taking a logarithm at the level of  partition functions.
For the pair of odd traces there is no difference, $K_{2m_1-1,2m_2-1} = C_{2m_1-1,2m_2-1}$
and for $q=1$ their Laplace transform in $N$ is given by the double $s$-expansion
of arctan in (\ref{rho}).
It is convenient to introduce additional parameter $\beta$ in denominator of this formula,
so that
\be
\sum_{N=0}^\infty \lambda^N \sum_{m_1\geq m_2\geq 1}
\frac{s_1^{2m_1-1}s_2^{2m_2-1}}{(2m_1-1)!!\,(2m_2-1)!!}
 \langle \tr X^{2m_1-1}\, \tr X^{2m_2-1}  \rangle_{q=1} =
\nn \\
= \left.\sum_{k=0} \frac{(-)^k (s_1s_2)^{2k+1}}{2k+1}
\frac{\lambda(1-\lambda)^{k-1}}
{\Big(1-\lambda - \beta(1+\lambda)(s_1^2+s_2^2)\Big)^{k+1}}\right|_{\beta=1}
= \nn \\
= \left. \sum_{k,j=0}   \frac{(-)^k}{2k+1} \frac{(k+j)!}{j!\cdot k!}
\frac{\beta^j\lambda(1+\lambda)^j}{(1-\lambda)^{2+j}} (s_1s_2)^{2k+1}(s_1^2+s_2^2)^j
\right|_{\beta=1}
\ee
Picking up appropriate powers of $s_1$ and $s_2$ we get for $m_1\geq m_2$
\be
\sum_{N=0}^\infty \lambda^N \sum_{m_1\geq m_2\geq 1}
\frac{ \langle \tr X^{2m_1-1}\ \tr X^{2m_2-1}  \rangle_{q=1}}{(2m_1-1)!!\,(2m_2-1)!!}
= \nn \\
= \left.\sum_{k=0}^{m_2-1}  \frac{(-)^k\beta^{m_1+m_2-2k-2}}{(2k+1)\cdot k!}
\frac{(m_1+m_2-2)!}{(m_1-k-1)!(m_2-k-1)!\cdot k!}
\frac{\lambda(1+\lambda)^{m_1+m_2-2k-2}}{(1-\lambda)^{m_1+m_2-2k}}
\right|_{\beta=1}
\ee

Thus for $q=1$  the polynomial in (\ref{odd2corr}) is
\be \label{classP}
\left.{\rm Pol}_{2m_2-2}(m_1|\lambda)\right|_{q=1} = \sum_{j=0}^{m_2-1} \frac{(-)^j}{2j+1} \cdot  
\frac{(2m_2-1)!!}{  j!\cdot (m2-j-1)!}
\cdot (1+\lambda)^{2(m_2-1-j)}\cdot (1-\lambda)^{2j}\cdot
  {\prod_{i=1}^{m_2-1}(m_1+i-j-1)}
\ee
$q$-deformation is relatively straightforward -- all factorials are changed for $q$-factorials
and powers of $(1\pm\lambda)$ to appropriate Pochhammer symbols.
Note only that  $m_1$ and $m_2$  are everywhere converted to $2m_1$ and $2m_2$,
and accordingly all factorials are double-factorials.
\be
%\boxed{
%\begin{array}{c}
P_{2m_2-2}(m_1|\lambda) =  \sum_{k=0}^{m_2-1} \frac{(-)^k}{[2k+1][2k]!!}
\frac{[2m_2-1]!!}{[2m_2-2-2k]!!}  \cdot
\frac{[2m_1+2m_2-4-2k]!!}{[2m_1-2-2k]!!}
\cdot \nn\\ \nn\\
\cdot
\prod_{i=1}^{2m_2-2-2k} (q^{2m_1-2m_2 +2i}+\lambda)\cdot
\prod_{i=2m_2-1-2k}^{2m_2-2} (q^{2m_1-2m_2 +2i}-\lambda)
%\end{array}
%}
\ee
The first two of these polynomials coincide with the examples in (\ref{odd2corr}),
but the structure is now clarified and inspired by (\ref{classP}).
Odd double-trace correlators are now provided by the last expression in (\ref{odd2corr}):\footnote{
We observe here an interesting phenomenon -- conversion of the naive ratio
with non-trivial switch in the numerator:
$$
\frac{\prod_{i=1}^j(q^{2i}+\lambda)}{\prod_{i=0}^{j+1}(1-q^{2i}\lambda)}
= \frac{\prod_{i=1}^j(q^{2i}+\lambda)\prod_{i=j+2}^{j_{\rm max}+1}(1-q^{2i}\lambda)}
{\prod_{i=0}^{j_{\rm max}+1}(1-q^{2i}\lambda)}
\ \longrightarrow \ 
\frac{\prod_{i=1}^j(q^{2i}+\lambda)\prod_{i=j+1}^{j_{\rm max}}(q^{2i}-\lambda)}
{\prod_{i=0}^{j_{\rm max}+1}(1 -q^{2i}\lambda)}
$$
Here $j=m_1+m_2-2-2k$ with $k\geq 0$ and $j_{\rm max} = m_1+m_2-2$.
Modulo a common power of $q$.
the switch is basically the change in one of the products $q^{2i}\longrightarrow q^{2-2i}$,
i.e. a kind of a twisted conjugation in the case of a unimodular $q=e^{i\pi \alpha}$.
}
\be \label{eq:two-point-hz-odd-case-generic}
\boxed{
\begin{array}{c}
\sum_{N=0}^\infty \lambda^N \frac{\left \langle \tr X^{2m_1-1}\, \tr X^{2m_2-1} \right \rangle_q}
{[2m_1-1]!!\cdot[2m_2-1]!!}
=   \\ \\
=    \sum_{k=0}^{m_2-1} \frac{(-)^k}{[2k+1][2k]!!}
 \cdot
\frac{{q^{m_1+m_2-1}}[2m_1+2m_2-4-2k]!!}{[2m_1-2-2k]!![2m_2-2-2k]!!}
%\cdot \\ \\
\cdot \frac{\lambda
\prod_{i=1}^{m_1+m_2-2-2k} (q^{2i}+\lambda)\cdot
\prod_{i=m_1+m_2-1-2k}^{m_1+m_2-2} (q^{ 2i}-\lambda)}
{\prod_{i=0}^{m_1+m_2-1}(1-q^{2i}\lambda)}
\end{array}
}
\ee

Irreducible even double-trace correlators are described by equally explicit, even if a bit lengthier, formula
\begin{align}
  {\rm  for}\ m_1\geq m_2 \ \ \ \ \ &\ 
  \sum_{N=0}^\infty \lambda^N \left \langle \left \langle\tr X^{2m_1}\, \tr X^{2m_2}
  \right \rangle\right \rangle_q
  := \sum_{N=0}^\infty \lambda^N \left(\left \langle  \tr X^{2m_1}\, \tr X^{2m_2} \right \rangle_q
  - \left \langle  \tr X^{2m_1}\  \right \rangle_q
  \left \langle  \ \tr X^{2m_2} \  \right \rangle_q    \right) \
  = \notag \\
  = & [2 m_1 - 1]!! [2 m_2 - 1]!! \lambda \left(\lambda + q^{2(m_1 + m_2)}\right)
  q^{m_1 - m_2 + 1} \frac{\prod_{i=1}^{m_1 - m_2} (\lambda + q^{2 i})}
  {\prod_{i=0}^{m_1 + m_2} (1 - \lambda q^{2 i})}
  \cdot \text{Pol}^{(\text{even})}_{2 m_2 - 2} (\lambda | m_1),
\end{align}
with
\begin{align}
  \label{eq:even-2-point-explicit-formula}
  \text{Pol}^{(\text{even})}_{2 n} (\lambda | m_1) = & \
  \sum_{j = 0}^{n - 1} \frac{[2 (2 n - 2 j)]}{[2 j + 1][2 j]!! [2 n - 2 j][2 n - 2 j]!!}
  \frac{[2 m_1 + 2 n - 2 j]!!}{[2 m_1 - 2 - 2 j]!!}
  \frac{(-1)^j}{2 }
  \text{p}_{2 n, 2 j} (\lambda |m_1) \\ \notag
  + & \ \sum_{j = 0}^{n - 2}
  \frac{\{q\}}{[2 j + 1][2 j]!! [2 n - 2 - 2 j]!!}
  \frac{[2 m_1 + 2 n - 2 j]!!}{[2 m_1 - 2 - 2 j]!!}
  \frac{(-1)^j}{2 }
  \text{p}_{2 n, 2 j + 1} (\lambda |m_1)
  \\ \notag
  + & \
  \frac{1}{[2 n + 1][2 n]!!}
  \frac{[2 m_1]!!}{[2 m_1 - 2 - 2 n]!!}
  (-1)^n
  \text{p}_{2 n, 2 n} (\lambda |m_1),
\end{align}
where $\{q\} = q - q^{-1}$ and the basis polynomials $\text{p}_{2 n, k}$ are similar to
the structures in the odd double-trace correlator:
\begin{align}
  \text{p}_{2 n, k} (\lambda |m_1) =
  \prod_{l=0}^{k-1} (\lambda - q^{2 m_1 + 2 n - 2 l})
  \prod_{l=k}^{2 n - 1} (\lambda + q^{2 m_1 + 2 n - 2 l})
\end{align}

In general, relation between even and odd cases is:
\be \label{eq:relation-between-even-and-odd-hz-two-points}
{\rm  for}\ m_1\geq m_2 \ \ \ \ \ 
\sum_{N=0}^\infty \lambda^N \left \langle \left \langle\tr X^{2m_1}\, \tr X^{2m_2}
\right \rangle\right \rangle_q
:= \sum_{N=0}^\infty \lambda^N \left(\left \langle  \tr X^{2m_1}\, \tr X^{2m_2} \right \rangle_q
 - \left \langle  \tr X^{2m_1}\  \right \rangle_q
 \left \langle  \ \tr X^{2m_2} \  \right \rangle_q    \right) \
= \nn \\
=  \frac{q^{2m_1+2m_2}+\lambda}{q^{2m_1+2-2m_2}+\lambda}
\sum_{j=0}^{m_2-1} \frac{[2m_1+2-2m_2+4j]\cdot [2m_1-1]!!\cdot [2m_2-1]!!}
{q^{2m_2-2j-1}\cdot[2m_1+2j+1]!!\cdot [2m_2-1-2j]!!}\cdot 
 \prod_{i=1}^j \frac{1+q^{2m_1-2m_2+4i-2}\lambda}{q^{2m_1 - 2m_2 +4i+2}+\lambda}
\cdot \nn \\
\cdot \sum_{N=0}^\infty \lambda^N \left \langle  \tr X^{2m_1+2j+1}\, \tr X^{2m_2-1-2j} \
\right \rangle_q \ \ \ \ \ \ \ \ \ \ 
\ee

Explicit interpretation of the r.h.s. as an action of a difference operator, which would look
as literal generalization of \eqref{rho}, will be given elsewhere, together with analysis of
the multi-trace case.
 
}

{\section{Towards $(q,t)$-Harer-Zagier formulas}\label{sec:qt-generalization}
  In this section we list our partial progress in understanding the $(q,t)$ generalization
  of the Harer-Zagier formulas.
  For the purposes of this section we define the $(q,t)$-deformation of Hermitian Gaussian matrix model (qtHGMM)
  by means of its Macdonald averages, as was done in \cite{paper:MPSh-on-qt-deformation-of-gaussian-mm}.
  This definition implies poles at $q=1$ (for $t$ fixed) and some other unexpected properties --
  which seem technically difficult to avoid, but do not have any clear conceptual explanation.

  With this definition the simplest 1-point correlator is
  \begin{align} \label{eq:trace-x2-qt}
    \left \langle \ \tr (X^2) \ \right \rangle_{q,t}:=
    &\left \langle \ \Mac_{[2]}-\frac{\{t\}(q+q^{-1})}{\{qt\}}\Mac_{[1,1]} \ \right \rangle_{q,t}=\nn\\
    &  = \frac{\{t^N\}}{\{q\}\{t\}}\cdot\left((q^2-1)t^{N-1} + \frac{2\{t^{N-1}\}}{t^2+1}\right),
  \end{align}
  where $\{x\} = x - x^{-1}$.
  We use notation with the trace, though in Macdonald case the proper notion of single-trace operators
  is not so obvious -- and this can affect the interpretation of eq.(\ref{eq:trace-x2-qt2}) below.

  The Laplace transform of (\ref{eq:trace-x2-qt}) does have reasonable poles,
  but the numerator is not a simple factorized expression
  \begin{align} \label{eq:trace-x2-qtl}
    \sum_{N=0}^\infty  \lambda^N\left \langle \ \tr (X^2) \ \right \rangle_{q,t}
    =  \frac{q\lambda}{(\lambda-1)(\lambda t^2-1)}
    \left(1-\frac{2\lambda(t^2-1)}{(\lambda-t^2)(q^2-1)}\right)
    = \nn \\
    = -\frac{q\lambda}{(\lambda-t^2)(\lambda-1)(\lambda t^2-1)}\left(
    \lambda+t^2 +\frac{2\lambda(t^2-q^2)}{q^2-1}\right)
  \end{align}

  Factorized is the answer for (so far mysterious) linear combination
  \begin{align} \label{eq:trace-x2-qt2}
    \boxed{
      \sum_{N=0}^\infty  \lambda^N\left \langle \ \frac{q^2-1}{t^2-1}\cdot\tr (X^2)
      -\frac{q^2-t^2}{t^2-1}\cdot (\tr X)^2 \ \right \rangle_{q,t}
      =    -\frac{q\lambda(\lambda+t^2)}{(\lambda-t^2)(\lambda-1)(\lambda t^2-1)}
    }
  \end{align}

  \bigskip

  Likewise, for the second non-trivial 1-point correlator
  \begin{align} \label{eq:trace-x4-qt}
    \scriptstyle
    \left \langle \ \tr (X^4) \ \right \rangle_{q,t} = & \nn\\
    \scriptstyle
    = & \ \scriptstyle \frac{\{t^N\}}{\{q\}^2\{t\}^4}\cdot\left(
    \frac{q^8t^8+q^8t^6-q^6t^8+q^8t^4+q^8t^2-q^2t^8+q^4t^4+t^8+q^6-t^6+q^4+q^2-t^2+1}{t^8}\cdot t^{3N}
    - \right.\nn \\ \scriptstyle & \ \scriptstyle \left.
    - \frac{q^6t^t+q^6t^4+q^6t^2+q^4t^4+q^2t^6+q^6+q^2t^4-2t^6+q^4+q^2*t^2+t^4+q^2+1}{t^6}\cdot t^N
    + 2t^{-N}+2t^{-3N}\right)
  \end{align}

  and its Laplace transform
  \begin{align} \label{eq:trace-x4-qtl}
    \scriptstyle
    \sum_{N=0}^\infty  \lambda^N\left \langle \ \tr (X^4) \ \right \rangle_{q,t}
    =  \frac{ q^2\lambda}{(\lambda- t^4 )(\lambda-1)(\lambda t^2-1)(\lambda t^4-1)}
    \left( t^4(q^4+q^2+1 )+\frac{\lambda  (A\lambda+B)}{ t^2(q^2-1)^2}\right)
    =\nn \\
    \scriptstyle
    = \frac{ q^2\lambda}{(\lambda- t^4 )(\lambda-1)(\lambda t^2-1)(\lambda t^4-1)}
    \left( \frac{q^4+q^2+1}{t^2}(\lambda+t^4)(\lambda+t^2)
    +\frac{(q^2-t^2)\lambda  (\tilde A\lambda+\tilde B)}{ t^2(q^2-1)^2}\right)
  \end{align}

  In the absence of full understanding, let us speculate a little bit.
  We know, that the basis (in the space of symmetric functions) of Macdonald polynomials
  has simple factorized averages, as functions of $q$, $t$ and $Q = t^N$.
  At the same time, when we take the Laplace transform of the Macdonald averages, this simple
  factorizability is lost, and instead some (yet unknown) \textit{Harer-Zagier} symmetric functions
  have simple factorized averages.

  In $q = 1$, and even in $q \neq 1, t = q$ case the one-point (symmetric) Harer-Zagier functions
  were just simple power sums. However, the generic $(q,t)$ case seems to tell us, that
  in general these functions are a little bit more complicated.

  Perhaps, one can, indeed, look for a basis in the space of symmetric functions that
  have nicely factorized Laplace-transformed averages and (together with some form of orthogonality)
  this will be enough to fix them. The work in continuing in these directions.
}

{\section{Towards q-Wick theorem and q-genus expansion}\label{sec:q-genus-expansion}
  %% ### vv Here we describe the partial successes of the genus expansion
  %% ###    and the hopes

  At the $q = 1$ any trace product correlator (due to the Wick theorem) is a sum
  over the fat (ribbon) graphs, where the valencies of the graph's vertices are equal
  to the exponents of the matrix under the trace. The weight of each particular fat
  graph is (up to an overall normalization factor)
  $$N^{\text{Euler characteristic}}$$

  For instance the average of $\tr (X^4)$ is equal to the following sum
  \begin{align}
    \begin{picture}(300,70)(0,-70)
      \put(0,-20){
        \put(0,0){$\left\langle \tr (X^4) \right\rangle$}
        \put(50,0){$=$}
        \put(70,0){$N \cdot \Big($}
        \put(95,0){$N^2$}
        \put(125,0){$+$}
        \put(150,0){$N^2$}
        \put(185,0){$+$}
        \put(220,0){$N^0$}
        \put(260,0){$\Big)$}
      }
      %% ### vv The tr X^4
      \put(15,-50){
        \put(0,0){\line(-1,0){10}}
        \put(0,0){\line(0,1){10}}
        \put(5,0){
          \put(0,0){\line(1,0){10}}
          \put(0,0){\line(0,1){10}}
        }
        \put(5,-5){
          \put(0,0){\line(1,0){10}}
          \put(0,0){\line(0,-1){10}}
        }
        \put(0,-5){
          \put(0,0){\line(-1,0){10}}
          \put(0,0){\line(0,-1){10}}
        }
      }
      %% ### vv The first sphere gluing
      \put(100,-50){
        \put(0,0){\line(-1,0){10}}
        \put(0,0){\line(0,1){10}}
        \put(5,0){
          \put(0,0){\line(1,0){10}}
          \put(0,0){\line(0,1){10}}
        }
        \put(5,-5){
          \put(0,0){\line(1,0){10}}
          \put(0,0){\line(0,-1){10}}
        }
        \put(0,-5){
          \put(0,0){\line(-1,0){10}}
          \put(0,0){\line(0,-1){10}}
        }
        \put(-10,0){\qbezier(0,0)(-5,5)(0,10)}
        \put(-10,-5){\qbezier(0,0)(-13,10)(0,20)}
        \put(0,10){\qbezier(0,0)(-5,5)(-10,0)}
        \put(5,10){\qbezier(0,0)(-10,13)(-20,0)}
        \put(15,-5){\qbezier(0,0)(5,-5)(0,-10)}
        \put(5,-15){\qbezier(0,0)(5,-5)(10,0)}
        \put(15,0){\qbezier(0,0)(13,-10)(0,-20)}
        \put(0,-15){\qbezier(0,0)(10,-13)(20,0)}
      }
      %% ### vv The second sphere gluing
      \put(150,-50){
        \put(0,0){\line(-1,0){10}}
        \put(0,0){\line(0,1){10}}
        \put(5,0){
          \put(0,0){\line(1,0){10}}
          \put(0,0){\line(0,1){10}}
        }
        \put(5,-5){
          \put(0,0){\line(1,0){10}}
          \put(0,0){\line(0,-1){10}}
        }
        \put(0,-5){
          \put(0,0){\line(-1,0){10}}
          \put(0,0){\line(0,-1){10}}
        }
        \put(5,0){
          \put(10,0){\qbezier(0,0)(5,5)(0,10)}
          \put(10,-5){\qbezier(0,0)(13,10)(0,20)}
          \put(0,10){\qbezier(0,0)(5,5)(10,0)}
          \put(-5,10){\qbezier(0,0)(10,13)(20,0)}
          \put(-15,-5){\qbezier(0,0)(-5,-5)(0,-10)}
          \put(-5,-15){\qbezier(0,0)(-5,-5)(-10,0)}
          \put(-15,0){\qbezier(0,0)(-13,-10)(0,-20)}
          \put(0,-15){\qbezier(0,0)(-10,-13)(-20,0)}
        }
      }
      %% ### vv The torus gluing
      \put(220,-50){
        \put(0,0){\line(-1,0){10}}
        \put(0,0){\line(0,1){10}}
        \put(5,0){
          \put(0,0){\line(1,0){10}}
          \put(0,0){\line(0,1){10}}
        }
        \put(5,-5){
          \put(0,0){\line(1,0){10}}
          \put(0,0){\line(0,-1){10}}
        }
        \put(0,-5){
          \put(0,0){\line(-1,0){10}}
          \put(0,0){\line(0,-1){10}}
        }
        %% ### vv One band
        \put(-10,0){\qbezier(0,0)(-10,10)(0,15)}
        \put(-10,15){\qbezier(0,0)(10,5)(25,0)}
        \put(15,15){\qbezier(0,0)(10,-5)(0,-15)}
        \put(-10,-5){\qbezier(0,0)(-18,15)(0,25)}
        \put(-10,20){\qbezier(0,0)(12,6)(25,0)}
        \put(15,20){\qbezier(0,0)(18,-10)(0,-25)}
        %% ### vv Another band
        \put(5,10){\qbezier(0,0)(2,3)(5,0)}
        \put(10,10){\qbezier(0,0)(3,-5)(2,-10)}
        \put(0,10){\qbezier(0,0)(6,10)(15,0)}
        \put(15,10){\qbezier(0,0)(3,-5)(2,-8)}
        \put(0,-5){
          \put(5,-10){\qbezier(0,0)(2,-3)(5,0)}
          \put(10,-10){\qbezier(0,0)(3,5)(2,10)}
          \put(0,-10){\qbezier(0,0)(6,-10)(15,0)}
          \put(15,-10){\qbezier(0,0)(3,5)(3,13)}
        }
      }
    \end{picture}
  \end{align}

  In other words, this sum is a sum over gluings of Riemann surfaces with marked points
  from polygons, each polygon corresponding to a vertex of a fat graph (and to a trace of the matrix)
  and thus having as many edges as the corresponding exponent. Continuing the example, there are the following gluings
  \begin{align}
    \begin{picture}(300,40)(0,-40)
      \put(90,0){
        \put(0,0){\line(1,0){30}}
        \put(30,0){\line(0,-1){30}}
        \put(30,-30){\line(-1,0){30}}
        \put(0,-30){\line(0,1){30}}
        %% ### vv Marks
        \thicklines
        \put(0,-15){\put(-1,-1){\qbezier(0,0)(1,1)(2,2)}}
        \put(15,0){\put(-1,-1){\qbezier(0,0)(1,1)(2,2)}}
        \put(30,-15){\put(-1,1){\qbezier(0,0)(1,1)(2,2)}}
        \put(30,-15){\put(-1,-1){\qbezier(0,0)(1,1)(2,2)}}
        \put(15,-30){\put(-2,-1){\qbezier(0,0)(1,1)(2,2)}}
        \put(15,-30){\put(0,-1){\qbezier(0,0)(1,1)(2,2)}}
      }
      \put(140,0){
        \put(0,0){\line(1,0){30}}
        \put(30,0){\line(0,-1){30}}
        \put(30,-30){\line(-1,0){30}}
        \put(0,-30){\line(0,1){30}}
        %% ### vv Marks
        \thicklines
        \put(0,-15){\put(-1,-1){\qbezier(0,0)(1,1)(2,2)}}
        \put(15,0){\put(0,-1){\qbezier(0,0)(1,1)(2,2)}}
        \put(15,0){\put(-2,-1){\qbezier(0,0)(1,1)(2,2)}}
        \put(30,-15){\put(-1,1){\qbezier(0,0)(1,1)(2,2)}}
        \put(30,-15){\put(-1,-1){\qbezier(0,0)(1,1)(2,2)}}
        \put(15,-30){\put(-1,-1){\qbezier(0,0)(1,1)(2,2)}}
      }
      \put(210,0){
        \put(0,0){\line(1,0){30}}
        \put(30,0){\line(0,-1){30}}
        \put(30,-30){\line(-1,0){30}}
        \put(0,-30){\line(0,1){30}}
        %% ### vv Marks
        \thicklines
        \put(0,-15){\put(-1,-1){\qbezier(0,0)(1,1)(2,2)}}
        \put(15,0){\put(0,-1){\qbezier(0,0)(1,1)(2,2)}}
        \put(15,0){\put(-2,-1){\qbezier(0,0)(1,1)(2,2)}}
        \put(30,-15){\put(-1,1){\qbezier(0,0)(1,1)(2,2)}}
        \put(15,-30){\put(-2,-1){\qbezier(0,0)(1,1)(2,2)}}
        \put(15,-30){\put(0,-1){\qbezier(0,0)(1,1)(2,2)}}
      }
    \end{picture}
  \end{align}
  The factor of $N$ then comes from the additional sum over the labellings
  of the gluings' distinct vertices with numbers $1 \ ..\ N$.
  In the example, the first and the second gluings have three distinct vertices,
  resulting in the factor $N \cdot N^2$, while the third gluing has one distinct
  vertex and has weight $N \cdot N^0$.

  After going to $q \neq 1$, for each concrete integer $N$ any trace product correlator
  becomes a Laurent polynomial in $q$ with non-negative coefficients.
  For instance, for $\tr (X^4)$ at first few $N$
  \begin{align}
    \left\langle \tr (X^4) \right\rangle_q \Bigg{|}_{N=1} = & \ q^2 + q^4 + q^6
    \\ \notag
    \left\langle \tr (X^4) \right\rangle_q \Bigg{|}_{N=2} = & \ 2 q^{-2} + 3 q^0 + 4 q^2 + 3 q^4 + 3 q^6 + 2 q^8 + q^{10}
    \\ \notag
    \left\langle \tr (X^4) \right\rangle_q \Bigg{|}_{N=3} = & \ 2 q^{-6} + 4 q^{-4} + 6 q^{-2} + 7 q^0 + 9 q^2 + 9 q^4 + 8 q^6 + 5 q^8 + 4 q^{10} + 2 q^{12} + q^{14}
  \end{align}

  One can easily check with help of computer experiments (both for one-point and multi-point correlators) that
  \begin{align} \label{eq:equality-of-number-of-monomials}
    \boxed{
      \left\{
      \begin{array}{l}
        \# \text{ of $q$-monomials} \\
        \text{in a polynomial,} \\
        \text{with multiplicities}
      \end{array}
      \right\}
      =
      \left\{
      \begin{array}{c}
        \# \text{ of labeled} \\
        \text{gluings}
      \end{array}
      \right\}
    }
  \end{align}

  It is, therefore, tempting to conjecture that for $q \neq 1$ some formula
  for trace product correlators in terms of labeled polygon gluings exists,
  namely, that there is some recipe to assign a $q$-monomial to each labeled gluing.

  Such a formula would simultaneously provide a $q$-analogue of the Wick theorem,
  because it would assign a $q$-weight to (labelled) pairwise splitting of the sides
  of the polygons,
  and the $q$-analogue of genus expansion,
  because every Riemann surface, resulting from a gluing, naturally has a genus.

  Obtaining such formula requires careful study (because straightforward attempts do not succeed)
  and is postponed for future research. Here are the first two examples,
  which illustrate some of the occuring complications.
  Ignoring the labels, each of these correlators should have exactly one gluing
  \begin{align}
    \left\langle (\tr X)^2 \right\rangle = N \
    & \mathop{\rightarrow}_{q \neq 1} \left\langle (\tr X)^2 \right\rangle_q = q^{N} \frac{(q^{-N} - q^N)}{(q^{-1} - q)}
    \\ \notag
    \left\langle (\tr X^2) \right\rangle = N^2 \
    & \mathop{\rightarrow}_{q \neq 1} \left\langle (\tr X^2) \right\rangle_q =
    \frac{(q^{-N} - q^N)}{(q^{-1} - q)}
    \frac{\left(q^{2 N} q^2 + q^{2 N} q^{-2} - 2 \right)}
    {q^N (q + q^{-1})}
  \end{align}

  One sees that two kinds of phenomena occur. First, spurious $q^N$ framing factors (vanishing at $q \rightarrow 1$) appear.
  Second, the deformation is clearly more complicated than the naive quantization $N \rightarrow [N]$.
  While our q-Harer-Zagier formulas \eqref{eq:explicit-residues1}, \eqref{eq:explicit-residues2},
  together with the observation \eqref{eq:equality-of-number-of-monomials} do provide
  evidence that the correct quantization prescription can be found, this is neither immediate, nor obvious
  and, once available, will undoubtedly lead to better understanding of the relevant enumerative geometric structures.
}

{\section{Conclusion}\label{sec:conclusion}
  To conclude, we derived an {\bf explicit q-deformation of Harer-Zagier formula}
  for the single-trace operator averages in Gaussian Hermitian model,
  which looks somewhat involved for particular matrix zises $N$,
  but actually becomes {\bf as simple as it was at $q=1$ after Laplace transform in $N$}:
  see Eq.~\eqref{eq:q-harer-zagier-concise}.
  We also conjectured, based on intuition from $q = 1$ case, equally explicit
  formulas for q-deformation of connected double-trace correlators,
  see Eqs.~\eqref{eq:two-point-hz-odd-case-generic}-\eqref{eq:relation-between-even-and-odd-hz-two-points}.

  The derivation can be based both on Jackson-integral version of the eigenvalue integral
  and on the super-integrability property of matrix models
  \cite{paper:IMM-complete-solution-to-gaussian-tensor-model,
    paper:MM-correlators-in-tensor-models-from-character-calculus},
  which is directly generalizable in many directions —
  from $q,t$-deformations to tensor models.
  However, four immediate generalizations of the remarkable results
  \eqref{eq:q-harer-zagier-concise} and
  \eqref{eq:two-point-hz-odd-case-generic}-\eqref{eq:relation-between-even-and-odd-hz-two-points}
  —
  to multi-trace correlators, to $t$ deformation, to q-genus expansion and to q-Wick theorem --
  appear somewhat problematic or at least less straightforward.

  Resolution of emerging problems should shed additional light on the structure
  of matrix-model deformations and their relation to representation theory of
  double loop algebras from the DIM family.
  This, in turn, is important for understaniding the relation between super-integrability
  (roughly, the existence of superficially large number of integrals of motion)
  and the more familiar KP/Toda/Hirota integrability in situations where the former one
  is obviously preserved, but the latter one needs serious and still unknown modification.
}

{\section*{Acknowledgments}
  This work was supported by the Russian Science Foundation (Grant No.20-12-00195).
}

\bibliographystyle{mpg}
\bibliography{references_q-harer-zagier}

\begin{thebibliography}{10}
\newcommand{\enquote}[1]{``#1''}
\providecommand{\url}[1]{\texttt{#1}}
\providecommand{\urlprefix}{ }
\providecommand{\eprint}[2][]{\url{#2}}

\bibitem{book:M-random-matrices}
Mehta, M.~L.
\newblock Random matrices.
\newblock Elsevier (2004).

\bibitem{paper:MM-exact-equation-for-the-loop-average-in-multicolor-qcd}
Makeenko, Y.~M. and Migdal, A.~A.
\newblock \enquote{Exact equation for the loop average in multicolor qcd}.
\newblock In: \enquote{The Large N Expansion In Quantum Field Theory And
  Statistical Physics: From Spin Systems to 2-Dimensional Gravity}, pp.
  227--229. World Scientific (1993).

\bibitem{paper:IZ-the-planar-approximation-II}
Itzykson, C. and Zuber, J.-B.
\newblock \enquote{The planar approximation. {II}}.
\newblock Journal of Mathematical Physics, vol.~21(3):pp. 411--421 (1980).

\bibitem{paper:K-the-appearance-of-matter-fields}
Kazakov, V.
\newblock \enquote{The appearance of matter fields from quantum fluctuations of
  2{D}-gravity}.
\newblock Modern Physics Letters A, vol.~4(22):pp. 2125--2139 (1989).

\bibitem{paper:KM-induced-qcs-at-large-n}
Kazakov, V. and Migdal, A.
\newblock \enquote{Induced {QCD} at large {N}, preprint {PUPT}-1322}.
\newblock Tech. rep., LPTENS-92/15, May (1992).

\bibitem{paper:M-integrability-and-matrix-models}
Morozov, A.~Y.
\newblock \enquote{Integrability and matrix models}.
\newblock Physics-Uspekhi, vol.~37(1):p.~1 (1994).
\newblock \href{http://arxiv.org/abs/hep-th/9303139}{{\ttfamily
  arXiv:hep-th/9303139 [hep-th]}}.

\bibitem{paper:M-matrix-models-as-integrable-systems}
Morozov, A.
\newblock \enquote{Matrix models as integrable systems}.
\newblock In: \enquote{Particles and fields}, pp. 127--210. Springer (1999).
\newblock \href{http://arxiv.org/abs/hep-th/9502091}{{\ttfamily
  arXiv:hep-th/9502091 [hep-th]}}.

\bibitem{paper:M-quantum-deformations-of-tau-functions}
Mironov, A.
\newblock \enquote{Quantum deformations of t-functions, bilinear identities and
  representation theory}.
\newblock Symmetries and Integrability of Difference Equations, vol.~9:pp.
  219--2 (1996).
\newblock \href{http://arxiv.org/abs/hep-th/9409190}{{\ttfamily
  arXiv:hep-th/9409190 [hep-th]}}.

\bibitem{paper:MMS-matrix-model-conjecture-for-exact-BS-periods}
Mironov, A., Morozov, A. and Shakirov, S.
\newblock \enquote{Matrix model conjecture for exact {BS} periods and
  {N}ekrasov functions}.
\newblock Journal of High Energy Physics, vol.
  2010(2)\newline\urlprefix\url{DOI:10.1007/jhep02(2010)030}
 (2010).
\newblock \href{http://arxiv.org/abs/0911.5721}{{\ttfamily arXiv:0911.5721
  [hep-th]}}.

\bibitem{paper:MMS-conformal-blocks-as-df-integral-discriminants}
Mironov, A., Morozov, A. and Shakirov, S.
\newblock \enquote{Conformal blocks as {D}otsenko–{F}ateev integral
  discriminants}.
\newblock International Journal of Modern Physics A, vol.~25(16):p.
  3173–3207\newline\urlprefix\url{DOI:10.1142/s0217751x10049141}
 (2010).
\newblock \href{http://arxiv.org/abs/1001.0563}{{\ttfamily arXiv:1001.0563
  [hep-th]}}.

\bibitem{paper:MMS-on-the-df-representation-of-toric}
Mironov, A., Morozov, A. and Shakirov, S.
\newblock \enquote{On the {“Dotsenko–Fateev”} representation of the toric
  conformal blocks}.
\newblock Journal of Physics A: Mathematical and Theoretical, vol.~44(8):p.
  085401\newline\urlprefix\url{DOI:10.1088/1751-8113/44/8/085401}
 (2011).
\newblock \href{http://arxiv.org/abs/1010.1734}{{\ttfamily arXiv:1010.1734
  [hep-th]}}.

\bibitem{paper:MMS-towards-a-proof-of-agt-mm}
Mironov, A., Morozov, A. and Shakirov, S.
\newblock \enquote{Towards a proof of {AGT} conjecture by methods of matrix
  models}.
\newblock International Journal of Modern Physics A, vol.~27(01):p.
  1230001\newline\urlprefix\url{DOI:10.1142/s0217751x12300013}
 (2012).
\newblock \href{http://arxiv.org/abs/1011.5629}{{\ttfamily arXiv:1011.5629
  [hep-th]}}.

\bibitem{paper:MMS-a-direct-proof-of-agt-at-beta-1}
Mironov, A., Morozov, A. and Shakirov, S.
\newblock \enquote{A direct proof of {AGT} conjecture at beta = 1}.
\newblock Journal of High Energy Physics, vol.
  2011(2)\newline\urlprefix\url{DOI:10.1007/jhep02(2011)067}
 (2011).
\newblock \href{http://arxiv.org/abs/1012.3137}{{\ttfamily arXiv:1012.3137
  [hep-th]}}.

\bibitem{paper:MMS-proving-agt-as-hs-duality}
Mironov, A., Morozov, A., Shakirov, S. and Smirnov, A.
\newblock \enquote{Proving {AGT} conjecture as {HS} duality: Extension to five
  dimensions}.
\newblock Nuclear Physics B, vol. 855(1):p.
  128–151\newline\urlprefix\url{DOI:10.1016/j.nuclphysb.2011.09.021}
 (2012).
\newblock \href{http://arxiv.org/abs/1105.0948}{{\ttfamily arXiv:1105.0948
  [hep-th]}}.

\bibitem{paper:IO-method-of-generating-q-expansion-coefficients}
Itoyama, H. and Oota, T.
\newblock \enquote{Method of generating q-expansion coefficients for conformal
  block and nekrasov function by beta-deformed matrix model}.
\newblock Nuclear Physics B, vol. 838(3):p.
  298–330\newline\urlprefix\url{DOI:10.1016/j.nuclphysb.2010.05.002}
 (2010).
\newblock \href{http://arxiv.org/abs/1003.2929}{{\ttfamily arXiv:1003.2929
  [hep-th]}}.

\bibitem{paper:ZP-localization-review}
Pestun, V., Zabzine, M., Benini, F., Dimofte, T., Dumitrescu, T.~T., Hosomichi,
  K., Kim, S., Lee, K., Le~Floch, B., Marino, M. et~al.
\newblock \enquote{Localization techniques in quantum field theories}.
\newblock Journal of Physics A: Mathematical and Theoretical, vol.~50(44):p.
  440301 (2017).
\newblock \href{http://arxiv.org/abs/1608.02952}{{\ttfamily arXiv:1608.02952
  [hep-th]}}.

\bibitem{paper:LPSZ-solving-q-virasoro-constraints}
Lodin, R., Popolitov, A., Shakirov, S. and Zabzine, M.
\newblock \enquote{Solving q-{V}irasoro constraints}.
\newblock Letters in Mathematical Physics, vol. 110(1):pp. 179--210 (2020).
\newblock \href{http://arxiv.org/abs/1810.00761}{{\ttfamily arXiv:1810.00761
  [hep-th]}}.

\bibitem{paper:CLPZ-exact-susy-wilson-loops-on-s3}
Cassia, L., Lodin, R., Popolitov, A. and Zabzine, M.
\newblock \enquote{Exact {SUSY} {W}ilson loops on {S3} from q-{V}irasoro
  constraints}.
\newblock Journal of High Energy Physics, vol. 2019(12):pp. 1--30 (2019).
\newblock \href{http://arxiv.org/abs/1909.10352}{{\ttfamily arXiv:1909.10352
  [hep-th]}}.

\bibitem{paper:CLZ-on-matrix-models-and-their-q-deformations}
Cassia, L., Lodin, R. and Zabzine, M.
\newblock \enquote{On matrix models and their $q$-deformations} (2020).
\newblock \href{http://arxiv.org/abs/2007.10354}{{\ttfamily arXiv:2007.10354
  [hep-th]}}.

\bibitem{paper:MNZ-5d-sym-and-ads7}
Minahan, J.~A., Nedelin, A. and Zabzine, M.
\newblock \enquote{5d super {Y}ang–{M}ills theory and the correspondence to
  {A}d{S}7/{CFT}6}.
\newblock Journal of Physics A: Mathematical and Theoretical, vol.~46(35):p.
  355401\newline\urlprefix\url{DOI:10.1088/1751-8113/46/35/355401}
 (2013).
\newblock \href{http://arxiv.org/abs/1304.1016}{{\ttfamily arXiv:1304.1016
  [hep-th]}}.

\bibitem{paper:NZ-q-virasoro-constraints-in-matrix-models}
Nedelin, A. and Zabzine, M.
\newblock \enquote{q-{V}irasoro constraints in matrix models}.
\newblock Journal of High Energy Physics, vol.
  2017(3)\newline\urlprefix\url{DOI:10.1007/jhep03(2017)098}
 (2017).
\newblock \href{http://arxiv.org/abs/1511.03471}{{\ttfamily arXiv:1511.03471
  [hep-th]}}.

\bibitem{paper:NNZ-q-virasoro-modular-double-and-3d-partition-functions}
Nedelin, A., Nieri, F. and Zabzine, M.
\newblock \enquote{q-{V}irasoro modular double and 3d partition functions}.
\newblock Communications in Mathematical Physics, vol. 353(3):p.
  1059–1102\newline\urlprefix\url{DOI:10.1007/s00220-017-2882-1}
 (2017).
\newblock \href{http://arxiv.org/abs/1605.07029}{{\ttfamily arXiv:1605.07029
  [hep-th]}}.

\bibitem{paper:MM-virasoro-constraints-for-kh-partition-function}
Mironov, A. and Morozov, A.
\newblock \enquote{Virasoro constraints for {K}ontsevich-{H}urwitz partition
  function}.
\newblock Journal of High Energy Physics, vol. 2009(02):p.
  024–024\newline\urlprefix\url{DOI:10.1088/1126-6708/2009/02/024}
 (2009).
\newblock \href{http://arxiv.org/abs/0807.2843}{{\ttfamily arXiv:0807.2843
  [hep-th]}}.

\bibitem{paper:AMMN-on-kp-integrable-hurwitz-functions}
Alexandrov, A., Mironov, A., Morozov, A. and Natanzon, S.
\newblock \enquote{On {KP}-integrable {H}urwitz functions}.
\newblock Journal of High Energy Physics, vol.
  2014(11)\newline\urlprefix\url{DOI:10.1007/jhep11(2014)080}
 (2014).
\newblock \href{http://arxiv.org/abs/1405.1395}{{\ttfamily arXiv:1405.1395
  [hep-th]}}.

\bibitem{paper:KL-combinatorial-solutions-to-integrable-hierarchies}
Kazarian, M.~E. and Lando, S.~K.
\newblock \enquote{Combinatorial solutions to integrable hierarchies}.
\newblock Russian Mathematical Surveys, vol.~70(3):p.
  453–482\newline\urlprefix\url{DOI:10.1070/rm2015v070n03abeh004952}
 (2015).
\newblock \href{http://arxiv.org/abs/1512.07172}{{\ttfamily arXiv:1512.07172
  [math.CO]}}.

\bibitem{paper:DBOPS-combinatorics-of-loop-equations-for-branched-covers-on-sphere}
Dunin-Barkowski, P., Orantin, N., Popolitov, A. and Shadrin, S.
\newblock \enquote{Combinatorics of loop equations for branched covers of
  sphere}.
\newblock International Mathematics Research Notices, vol. 2018(18):p.
  5638–5662\newline\urlprefix\url{DOI:10.1093/imrn/rnx047}
 (2017).
\newblock \href{http://arxiv.org/abs/1412.1698}{{\ttfamily arXiv:1412.1698
  [math-ph]}}.

\bibitem{paper:KPS-topological-recursion-for-monotone-orbifold}
Kramer, R., Popolitov, A. and Shadrin, S.
\newblock \enquote{Topological recursion for monotone orbifold {H}urwitz
  numbers: a proof of the {D}o-{K}arev conjecture} (2019).
\newblock \href{http://arxiv.org/abs/1909.02302}{{\ttfamily arXiv:1909.02302
  [math.AG]}}.

\bibitem{paper:DBKPS-cut-and-join-for-monotone}
Dunin-Barkowski, P., Kramer, R., Popolitov, A. and Shadrin, S.
\newblock \enquote{Cut-and-join equation for monotone {H}urwitz numbers
  revisited}.
\newblock Journal of Geometry and Physics, vol. 137:p.
  1–6\newline\urlprefix\url{DOI:10.1016/j.geomphys.2018.11.010}
 (2019).
\newblock \href{http://arxiv.org/abs/1807.04197}{{\ttfamily arXiv:1807.04197
  [math.AG]}}.

\bibitem{paper:K-recursion-for-masur-veech}
Kazarian, M.
\newblock \enquote{Recursion for {M}asur-{V}eech volumes of moduli spaces of
  quadratic differentials} (2019).
\newblock \href{http://arxiv.org/abs/1912.10422}{{\ttfamily arXiv:1912.10422
  [math-ph]}}.

\bibitem{paper:CKL-polynomial-graph-invariants-and-the-kp-hierarchy}
Chmutov, S., Kazarian, M. and Lando, S.
\newblock \enquote{Polynomial graph invariants and the {KP} hierarchy}.
\newblock Selecta Mathematica,
  vol.~26(3)\newline\urlprefix\url{DOI:10.1007/s00029-020-00562-w}
 (2020).
\newblock \href{http://arxiv.org/abs/1803.09800}{{\ttfamily arXiv:1803.09800
  [math.CO]}}.

\bibitem{paper:KZ-rationality-in-map-and-hypermap}
Zograf, P. and Kazarian, M.
\newblock \enquote{Rationality in map and hypermap enumeration by genus}.
\newblock St. Petersburg Mathematical Journal, vol.~29(3):p.
  439–445\newline\urlprefix\url{DOI:10.1090/spmj/1501}
 (2018).
\newblock \href{http://arxiv.org/abs/1609.05493}{{\ttfamily arXiv:1609.05493
  [math.CO]}}.

\bibitem{paper:KZ-virasoro-constraints-and-topological-recursion-for-dessins}
Kazarian, M. and Zograf, P.
\newblock \enquote{Virasoro constraints and topological recursion for
  {G}rothendieck’s dessin counting}.
\newblock Letters in Mathematical Physics, vol. 105(8):p.
  1057–1084\newline\urlprefix\url{DOI:10.1007/s11005-015-0771-0}
 (2015).

\bibitem{paper:DBKOSS-polynomiality-of-hurwitz-numbers-bm-conj}
Dunin-Barkowski, P., Kazarian, M., Orantin, N., Shadrin, S. and Spitz, L.
\newblock \enquote{Polynomiality of hurwitz numbers, bouchard–mariño
  conjecture, and a new proof of the elsv formula}.
\newblock Advances in Mathematics, vol. 279:p.
  67–103\newline\urlprefix\url{DOI:10.1016/j.aim.2015.03.016}
 (2015).
\newblock \href{http://arxiv.org/abs/1307.4729}{{\ttfamily arXiv:1307.4729
  [math.AG]}}.

\bibitem{paper:A-matrix-model-for-stationary-sector-of-gw-of-p1}
Alexandrov, A.
\newblock \enquote{Matrix model for the stationary sector of gromov-witten
  theory of ${\bf p}^1$} (2020).
\newblock \href{http://arxiv.org/abs/2001.08556}{{\ttfamily arXiv:2001.08556
  [math-ph]}}.

\bibitem{paper:ABT-refined-open-intersection-numbers-and-the-kp-mm}
Alexandrov, A., Buryak, A. and Tessler, R.~J.
\newblock \enquote{Refined open intersection numbers and the kontsevich-penner
  matrix model}.
\newblock Journal of High Energy Physics, vol.
  2017(3)\newline\urlprefix\url{DOI:10.1007/jhep03(2017)123}
 (2017).
\newblock \href{http://arxiv.org/abs/1702.02319}{{\ttfamily arXiv:1702.02319
  [math-ph]}}.

\bibitem{paper:A-open-intersection-numbers-and-free-fields}
Alexandrov, A.
\newblock \enquote{Open intersection numbers and free fields}.
\newblock Nuclear Physics B, vol. 922:p.
  247–263\newline\urlprefix\url{DOI:10.1016/j.nuclphysb.2017.06.019}
 (2017).
\newblock \href{http://arxiv.org/abs/1606.06712}{{\ttfamily arXiv:1606.06712
  [math-ph]}}.

\bibitem{paper:DL-on-the-goulden-jackson-vakil-conjecture-for-double-hurwitz-numbers}
Do, N. and Lewański, D.
\newblock \enquote{On the goulden-jackson-vakil conjecture for double hurwitz
  numbers} (2020).
\newblock \href{http://arxiv.org/abs/2003.08043}{{\ttfamily arXiv:2003.08043
  [math.AG]}}.

\bibitem{paper:BDKLM-double-hurwitz-numbers-polynomiality-topological-recursion-and-intersection-theory}
Borot, G., Do, N., Karev, M., Lewański, D. and Moskovsky, E.
\newblock \enquote{Double hurwitz numbers: polynomiality, topological recursion
  and intersection theory} (2020).
\newblock \href{http://arxiv.org/abs/2002.00900}{{\ttfamily arXiv:2002.00900
  [math.AG]}}.

\bibitem{paper:AKMM-shiraishi-functor-and-non-kerov}
Awata, H., Kanno, H., Mironov, A. and Morozov, A.
\newblock \enquote{Shiraishi functor and non-{K}erov deformation of {M}acdonald
  polynomials} (2020).
\newblock \href{http://arxiv.org/abs/2002.12746}{{\ttfamily arXiv:2002.12746
  [hep-th]}}.

\bibitem{paper:MM-on-hamiltonians-for-kerov-functions}
Mironov, A. and Morozov, A.
\newblock \enquote{On hamiltonians for {K}erov functions}.
\newblock The European Physical Journal C,
  vol.~80(3)\newline\urlprefix\url{DOI:10.1140/epjc/s10052-020-7811-3}
 (2020).
\newblock \href{http://arxiv.org/abs/1908.05176}{{\ttfamily arXiv:1908.05176
  [hep-th]}}.

\bibitem{paper:MM-kerov-functions-for-composite-representations}
Mironov, A. and Morozov, A.
\newblock \enquote{Kerov functions for composite representations and
  {M}acdonald ideal}.
\newblock Nuclear Physics B, vol. 944:p.
  114641\newline\urlprefix\url{DOI:10.1016/j.nuclphysb.2019.114641}
 (2019).
\newblock \href{http://arxiv.org/abs/1903.00773}{{\ttfamily arXiv:1903.00773
  [hep-th]}}.

\bibitem{paper:MM-kerov-functions-revisited}
Mironov, A. and Morozov, A.
\newblock \enquote{Kerov functions revisited}.
\newblock Journal of Geometry and Physics, vol. 150:p.
  103608\newline\urlprefix\url{DOI:10.1016/j.geomphys.2020.103608}
 (2020).
\newblock \href{http://arxiv.org/abs/1811.01184}{{\ttfamily arXiv:1811.01184
  [hep-th]}}.

\bibitem{paper:OP-the-equivariant-gw-theory-of-p1}
Okounkov, A. and Pandharipande, R.
\newblock \enquote{The equivariant gromov–witten theory of p1}.
\newblock Annals of Mathematics, vol. 163(2):p.
  561–605\newline\urlprefix\url{DOI:10.4007/annals.2006.163.561}
 (2006).
\newblock \href{http://arxiv.org/abs/math/0207233}{{\ttfamily
  arXiv:math/0207233 [math.AG]}}.

\bibitem{paper:OP-gw-theory-hurwitz-theory-and-completed-cycles}
Okounkov, A. and Pandharipande, R.
\newblock \enquote{Gromov-witten theory, hurwitz theory, and completed cycles}
  (2002).
\newblock \href{http://arxiv.org/abs/math/0204305}{{\ttfamily
  arXiv:math/0204305 [math.AG]}}.

\bibitem{paper:OP-gw-theory-hurwitz-numbers-and-matrix-models}
Okounkov, A. and Pandharipande, R.
\newblock \enquote{Gromov-witten theory, hurwitz numbers, and matrix models}.
\newblock Proceedings of Symposia in Pure Mathematics, p.
  325–414\newline\urlprefix\url{DOI:10.1090/pspum/080.1/2483941}
 (2009).
\newblock \href{http://arxiv.org/abs/math/0101147}{{\ttfamily
  arXiv:math/0101147 [math.AG]}}.

\bibitem{paper:SKDMSK-robust-architecture-for-programmable-universal-unitaries}
Saygin, M., Kondratyev, I., Dyakonov, I., Mironov, S., Straupe, S. and Kulik,
  S.
\newblock \enquote{Robust architecture for programmable universal unitaries}.
\newblock Physical Review Letters, vol.
  124(1)\newline\urlprefix\url{DOI:10.1103/physrevlett.124.010501}
 (2020).
\newblock \href{http://arxiv.org/abs/1906.06748}{{\ttfamily arXiv:1906.06748
  [quant-ph]}}.

\bibitem{paper:KM-quantum-r-matrices-as-universal-qubit-gates}
Kolganov, N. and Morozov, A.
\newblock \enquote{Quantum {R}-matrices as universal qubit gates}.
\newblock JETP Letters, vol. 111(9):p.
  519–524\newline\urlprefix\url{DOI:10.1134/s0021364020090027}
 (2020).
\newblock \href{http://arxiv.org/abs/2004.07764}{{\ttfamily arXiv:2004.07764
  [hep-th]}}.

\bibitem{paper:AC-the-matrix-model-for-dessins-denfants}
Ambjørn, J. and Chekhov, L.
\newblock \enquote{The matrix model for dessins d’enfants}.
\newblock Annales de l’Institut Henri Poincaré D, vol.~1(3):p.
  337–361\newline\urlprefix\url{DOI:10.4171/aihpd/10}
 (2014).
\newblock \href{http://arxiv.org/abs/1404.4240}{{\ttfamily arXiv:1404.4240
  [math.AG]}}.

\bibitem{paper:HZ-the-euler-characteristic-of-the-moduli-space-of-curves}
Harer, J. and Zagier, D.
\newblock \enquote{The euler characteristic of the moduli space of curves}.
\newblock Inventiones mathematicae, vol.~85(3):pp. 457--485 (1986).

\bibitem{paper:MSh-exact-2-point-function-in-hmm}
Morozov, A. and Shakirov, S.
\newblock \enquote{Exact 2-point function in {H}ermitian matrix model}.
\newblock Journal of High Energy Physics, vol. 2009(12):p. 003 (2009).
\newblock \href{http://arxiv.org/abs/0906.0036}{{\ttfamily arXiv:0906.0036
  [hep-th]}}.

\bibitem{paper:NS-quantization-of-integrable-systems-and-four-dimensional-gauge-theories}
Nekrasov, N.~A. and Shatashvili, S.~L.
\newblock \enquote{Quantization of integrable systems and four dimensional
  gauge theories}.
\newblock XVIth International Congress on Mathematical
  Physics\newline\urlprefix\url{DOI:10.1142/9789814304634_0015}
 (2010).
\newblock \href{http://arxiv.org/abs/0908.4052}{{\ttfamily arXiv:0908.4052
  [hep-th]}}.

\bibitem{paper:MM-nekrasov-functions-and-exact-bs-integrals}
Mironov, A. and Morozov, A.
\newblock \enquote{Nekrasov functions and exact {B}ohr-{S}ommerfeld integrals}.
\newblock Journal of High Energy Physics, vol.
  2010(4)\newline\urlprefix\url{DOI:10.1007/jhep04(2010)040}
 (2010).
\newblock \href{http://arxiv.org/abs/0910.5670}{{\ttfamily arXiv:0910.5670
  [hep-th]}}.

\bibitem{paper:MM-nekrasov-functions-from-exact-bs-periods-sun}
Mironov, A. and Morozov, A.
\newblock \enquote{Nekrasov functions from exact {B}ohr–{S}ommerfeld periods:
  the case of {SU(N)}}.
\newblock Journal of Physics A: Mathematical and Theoretical, vol.~43(19):p.
  195401\newline\urlprefix\url{DOI:10.1088/1751-8113/43/19/195401}
 (2010).
\newblock \href{http://arxiv.org/abs/0911.2396}{{\ttfamily arXiv:0911.2396
  [hep-th]}}.

\bibitem{paper:MMM-on-agt-relations-with-surface-operator-insertion-and-a-stationary-limit-of-beta-ensembles}
Marshakov, A., Mironov, A. and Morozov, A.
\newblock \enquote{On {AGT} relations with surface operator insertion and a
  stationary limit of beta-ensembles}.
\newblock Journal of Geometry and Physics, vol.~61(7):p.
  1203–1222\newline\urlprefix\url{DOI:10.1016/j.geomphys.2011.01.012}
 (2011).
\newblock \href{http://arxiv.org/abs/1011.4491}{{\ttfamily arXiv:1011.4491
  [hep-th]}}.

\bibitem{paper:NRS-darboux-coordinates-yy-functional-and-gauge-theory}
Nekrasov, N., Rosly, A. and Shatashvili, S.
\newblock \enquote{Darboux coordinates, {Y}ang-{Y}ang functional, and gauge
  theory}.
\newblock Nuclear Physics B - Proceedings Supplements, vol. 216(1):p.
  69–93\newline\urlprefix\url{DOI:10.1016/j.nuclphysbps.2011.04.150}
 (2011).
\newblock \href{http://arxiv.org/abs/1103.3919}{{\ttfamily arXiv:1103.3919
  [hep-th]}}.

\bibitem{paper:K-non-perturbative-quantum-geometry}
Krefl, D.
\newblock \enquote{Non-perturbative quantum geometry}.
\newblock Journal of High Energy Physics, vol.
  2014(2)\newline\urlprefix\url{DOI:10.1007/jhep02(2014)084}
 (2014).
\newblock \href{http://arxiv.org/abs/1311.0584}{{\ttfamily arXiv:1311.0584
  [hep-th]}}.

\bibitem{paper:K-non-perturbative-quantum-geometry-ii}
Krefl, D.
\newblock \enquote{Non-perturbative quantum geometry {II}}.
\newblock Journal of High Energy Physics, vol.
  2014(12)\newline\urlprefix\url{DOI:10.1007/jhep12(2014)118}
 (2014).
\newblock \href{http://arxiv.org/abs/1410.7116}{{\ttfamily arXiv:1410.7116
  [hep-th]}}.

\bibitem{paper:GMM-wall-crossing-invariants-from-quantum-mechanics-to-knots}
Galakhov, D., Mironov, A. and Morozov, A.
\newblock \enquote{Wall-crossing invariants: from quantum mechanics to knots}.
\newblock Journal of Experimental and Theoretical Physics, vol. 120(3):p.
  549–577\newline\urlprefix\url{DOI:10.1134/s1063776115030206}
 (2015).
\newblock \href{http://arxiv.org/abs/1410.8482}{{\ttfamily arXiv:1410.8482
  [hep-th]}}.

\bibitem{paper:MSS-the-qsc-for-double-hn}
Mulase, M., Shadrin, S. and Spitz, L.
\newblock \enquote{The spectral curve and the {S}chrödinger equation of double
  {H}urwitz numbers and higher spin structures}.
\newblock Communications in Number Theory and Physics, vol.~7(1):p.
  125–143\newline\urlprefix\url{DOI:10.4310/cntp.2013.v7.n1.a4}
 (2013).
\newblock \href{http://arxiv.org/abs/1301.5580}{{\ttfamily arXiv:1301.5580
  [math.AG]}}.

\bibitem{paper:DBMNPS-quantum-spectral-curve-for-gw-of-cp1}
Dunin-Barkowski, P., Mulase, M., Norbury, P., Popolitov, A. and Shadrin, S.
\newblock \enquote{Quantum spectral curve for the {G}romov–{W}itten theory of
  the complex projective line}.
\newblock Journal für die reine und angewandte Mathematik (Crelles Journal),
  vol. 2017(726)\newline\urlprefix\url{DOI:10.1515/crelle-2014-0097}
 (2017).
\newblock \href{http://arxiv.org/abs/1312.5336}{{\ttfamily arXiv:1312.5336
  [math-ph]}}.

\bibitem{paper:GGM-non-perturbative-approaches-to-the-quantum-sw-curve}
Grassi, A., Gu, J. and Mariño, M.
\newblock \enquote{Non-perturbative approaches to the quantum seiberg-witten
  curve}.
\newblock Journal of High Energy Physics, vol.
  2020(7)\newline\urlprefix\url{DOI:10.1007/jhep07(2020)106}
 (2020).
\newblock \href{http://arxiv.org/abs/1908.07065}{{\ttfamily arXiv:1908.07065
  [hep-th]}}.

\bibitem{paper:EMR-resonances-and-pt-symmetry-in-quantum-curves}
Emery, Y., Mariño, M. and Ronzani, M.
\newblock \enquote{Resonances and pt symmetry in quantum curves}.
\newblock Journal of High Energy Physics, vol.
  2020(4)\newline\urlprefix\url{DOI:10.1007/jhep04(2020)150}
 (2020).
\newblock \href{http://arxiv.org/abs/1902.08606}{{\ttfamily arXiv:1902.08606
  [hep-th]}}.

\bibitem{paper:AIMMVY-correspondence-between-feynman-diagrams}
Amburg, N., Itoyama, H., Mironov, A., Morozov, A., Vasiliev, D. and Yoshioka,
  R.
\newblock \enquote{Correspondence between {F}eynman diagrams and operators in
  quantum field theory that emerges from tensor model}.
\newblock The European Physical Journal C,
  vol.~80(5)\newline\urlprefix\url{DOI:10.1140/epjc/s10052-020-8013-8}
 (2020).
\newblock \href{http://arxiv.org/abs/1911.10574}{{\ttfamily arXiv:1911.10574
  [hep-th]}}.

\bibitem{paper:IMM-complete-solution-to-gaussian-tensor-model}
Itoyama, H., Mironov, A. and Morozov, A.
\newblock \enquote{Complete solution to {G}aussian tensor model and its
  integrable properties}.
\newblock Physics Letters B, vol. 802:p.
  135237\newline\urlprefix\url{DOI:10.1016/j.physletb.2020.135237}
 (2020).
\newblock \href{http://arxiv.org/abs/1910.03261}{{\ttfamily arXiv:1910.03261
  [hep-th]}}.

\bibitem{paper:IMM-from-kronecker-to-tableau}
Itoyama, H., Mironov, A. and Morozov, A.
\newblock \enquote{From {K}ronecker to tableau pseudo-characters in tensor
  models}.
\newblock Physics Letters B, vol. 788:p.
  76–81\newline\urlprefix\url{DOI:10.1016/j.physletb.2018.11.008}
 (2019).
\newblock \href{http://arxiv.org/abs/1808.07783}{{\ttfamily arXiv:1808.07783
  [hep-th]}}.

\bibitem{paper:MM-on-the-complete-perturbative-solution-of-on-matrix-models}
Mironov, A. and Morozov, A.
\newblock \enquote{On the complete perturbative solution of one-matrix models}.
\newblock Physics Letters B, vol. 771:p.
  503–507\newline\urlprefix\url{DOI:10.1016/j.physletb.2017.05.094}
 (2017).
\newblock \href{http://arxiv.org/abs/1705.00976}{{\ttfamily arXiv:1705.00976
  [hep-th]}}.

\bibitem{paper:MM-correlators-in-tensor-models-from-character-calculus}
Mironov, A. and Morozov, A.
\newblock \enquote{Correlators in tensor models from character calculus}.
\newblock Physics Letters B, vol. 774:p.
  210–216\newline\urlprefix\url{DOI:10.1016/j.physletb.2017.09.063}
 (2017).
\newblock \href{http://arxiv.org/abs/1706.03667}{{\ttfamily arXiv:1706.03667
  [hep-th]}}.

\bibitem{paper:CHPS-orbifolds-and-exact-solutions}
Córdova, C., Heidenreich, B., Popolitov, A. and Shakirov, S.
\newblock \enquote{Orbifolds and exact solutions of strongly-coupled matrix
  models}.
\newblock Communications in Mathematical Physics, vol. 361(3):p.
  1235–1274\newline\urlprefix\url{DOI:10.1007/s00220-017-3072-x}
 (2018).
\newblock \href{http://arxiv.org/abs/1611.03142}{{\ttfamily arXiv:1611.03142
  [hep-th]}}.

\bibitem{paper:MPSh-on-qt-deformation-of-gaussian-mm}
Morozov, A., Popolitov, A. and Shakirov, S.
\newblock \enquote{On (q, t)-deformation of {G}aussian matrix model}.
\newblock Physics Letters B, vol. 784:pp. 342--344 (2018).
\newblock \href{http://arxiv.org/abs/1803.11401}{{\ttfamily arXiv:1803.11401
  [hep-th]}}.

\bibitem{paper:IMM-tensorial-generalization-of-characters}
Itoyama, H., Mironov, A. and Morozov, A.
\newblock \enquote{Tensorial generalization of characters}.
\newblock Journal of High Energy Physics, vol.
  2019(12)\newline\urlprefix\url{DOI:10.1007/jhep12(2019)127}
 (2019).
\newblock \href{http://arxiv.org/abs/1909.06921}{{\ttfamily arXiv:1909.06921
  [hep-th]}}.

\bibitem{paper:M-cauchy-formula-and-the-character-ring}
Morozov, A.
\newblock \enquote{Cauchy formula and the character ring}.
\newblock The European Physical Journal C,
  vol.~79(1)\newline\urlprefix\url{DOI:10.1140/epjc/s10052-019-6598-6}
 (2019).
\newblock \href{http://arxiv.org/abs/1812.03853}{{\ttfamily arXiv:1812.03853
  [hep-th]}}.

\bibitem{paper:BDGMMMRSS-distinguishing-mutant-knots}
Bishler, L., Dhara, S., Grigoryev, T., Mironov, A., Morozov, A., Morozov, A.,
  Ramadevi, P., Singh, V.~K. and Sleptsov, A.
\newblock \enquote{Distinguishing mutant knots} (2020).
\newblock \href{http://arxiv.org/abs/2007.12532}{{\ttfamily arXiv:2007.12532
  [hep-th]}}.

\bibitem{paper:MST-a-novel-symmetry-of-colored-homfly}
Mishnyakov, V., Sleptsov, A. and Tselousov, N.
\newblock \enquote{A novel symmetry of colored {HOMFLY} polynomials coming from
  $\mathfrak{sl}(n|m)$ superalgebras} (2020).
\newblock \href{http://arxiv.org/abs/2005.01188}{{\ttfamily arXiv:2005.01188
  [hep-th]}}.

\bibitem{paper:BDGMMMRSS-difference-of-mutant-knot-invariants}
Bishler, L., Dhara, S., Grigoryev, T., Mironov, A., Morozov, A., Morozov, A.,
  Ramadevi, P., Singh, V.~K. and Sleptsov, A.
\newblock \enquote{Difference of mutant knot invariants and their differential
  expansion}.
\newblock JETP Letters, vol. 111(9):p.
  494–499\newline\urlprefix\url{DOI:10.1134/s0021364020090015}
 (2020).
\newblock \href{http://arxiv.org/abs/2004.06598}{{\ttfamily arXiv:2004.06598
  [hep-th]}}.

\bibitem{paper:MST-a-new-symmetry-of-colored-alexander}
Mishnyakov, V., Sleptsov, A. and Tselousov, N.
\newblock \enquote{A new symmetry of the colored {A}lexander polynomial}
  (2020).
\newblock \href{http://arxiv.org/abs/2001.10596}{{\ttfamily arXiv:2001.10596
  [hep-th]}}.

\bibitem{paper:MM-sum-rules-for-characters-from-character-preservation-property}
Mironov, A. and Morozov, A.
\newblock \enquote{Sum rules for characters from character-preservation
  property of matrix models}.
\newblock Journal of High Energy Physics, vol. 2018(8):p. 163 (2018).

\bibitem{book:M-symmetric-functions-and-hall-polynomials}
Macdonald, I.~G.
\newblock Symmetric functions and {H}all polynomials.
\newblock Oxford university press (1998).

\bibitem{book:KC-quantum-calculus}
Kac, V. and Cheung, P.
\newblock Quantum calculus.
\newblock Springer Science \& Business Media (2001).

\bibitem{paper:MMN-complete-set-of-cut-and-join-operators-in-the-hk-theory}
Mironov, A.~D., Morozov, A.~Y. and Natanzon, S.~M.
\newblock \enquote{Complete set of cut-and-join operators in the
  hurwitz-kontsevich theory}.
\newblock Theoretical and Mathematical Physics, vol. 166(1):p.
  1–22\newline\urlprefix\url{DOI:10.1007/s11232-011-0001-6}
 (2011).
\newblock \href{http://arxiv.org/abs/0904.4227}{{\ttfamily arXiv:0904.4227
  [hep-th]}}.

\end{thebibliography}

\end{document}